\documentclass[prl,twocolumn,preprintnumbers,footinbib,tightenlines,superscriptaddress]{revtex4}

\pdfoutput=1
\usepackage[english]{babel}
\makeatletter\AtBeginDocument{\let\@elt\relax}\makeatother 
\usepackage{amsmath,amssymb,amsfonts, bm,bbm,slashed, subdepth}
\usepackage{graphicx}
\usepackage{xcolor}
\usepackage{hyperref}
\usepackage{cleveref}
\usepackage{enumerate}
\usepackage{epsfig, subfigure}
\usepackage{setspace}
\usepackage{booktabs, tabularx}
\usepackage{units}
\usepackage{placeins}
\usepackage{multirow}
\usepackage{mathtools}
\usepackage{lipsum,epsfig,dsfont}
\usepackage{soul}
\usepackage{natbib}

\allowdisplaybreaks

\begin{document}
\title{Excess Radiation from Axion-Photon Conversion}
\author{Andrea Addazi}
\affiliation{Center for Theoretical Physics, College of Physics, Sichuan University,
Chengdu, 610064, PR China}
\affiliation{INFN, Laboratori Nazionali di Frascati, Via E. Fermi 54, I-00044 Roma,
Italy}
\author{Salvatore Capozziello}
\affiliation{Dipartimento di Fisica ”E. Pancini”, Università di Napoli “Federico
II”,  Complesso  Universitario  di Monte S. Angelo, Edificio G, Via Cinthia, I-80126,
Napoli, Italy}
\affiliation{Istituto Nazionale di Fisica Nucleare, Sezione di Napoli, Napoli,
Italy}
\affiliation{Scuola Superiore Meridionale, Largo S. Marcellino 10, I-80138, Napoli,
Italy}
\author{Qingyu Gan}
\affiliation{Scuola Superiore Meridionale, Largo S. Marcellino 10, I-80138, Napoli,
Italy}
\affiliation{Istituto Nazionale di Fisica Nucleare, Sezione di Napoli, Napoli,
Italy}
\author{Gaetano Lambiase}
\affiliation{Dipartimento di Fisica “E.R. Caianiello”, Università di Salerno, I-84084 Fisciano (Sa), Italy}
\affiliation{INFN- Gruppo Collegato di Salerno, I-84084 Fisciano (Sa), Italy}
\author{Rome Samanta}
\affiliation{Scuola Superiore Meridionale, Largo S. Marcellino 10, I-80138, Napoli,
Italy}
\affiliation{Istituto Nazionale di Fisica Nucleare, Sezione di Napoli, Napoli,
Italy}
\begin{abstract}
Two notable anomalies in radio observations — the excess radiation in the Rayleigh-Jeans tail of the cosmic microwave background, revealed by ARCADE-2, and the twice-deeper absorption trough of the global 21cm line, identified by EDGES — remain unresolved. These phenomena may have a shared origin, as the enhancement of the 21cm absorption trough could arise from excess heating. 
We investigate this scenario through the framework of axion-like particles (ALPs), showing that the resonant conversion of ALPs into photons can produce a photon abundance sufficient to resolve both anomalies simultaneously. 
Our model naturally explains the observed radio excess between 0.4 and 10GHz while also enhances the 21cm absorption feature at 78MHz. Furthermore, it predicts a novel power-law scaling of the radio spectrum above 0.5GHz and an additional absorption trough below 30MHz, which could be verified through cross-detection in upcoming experiments.
\end{abstract}

\date{\today}

\maketitle

%

\textbf{\emph{Introduction.}} The cosmic microwave background (CMB) at frequencies above 60 GHz stands as one of the most extensively studied electromagnetic signals in cosmology, supported  by numerous observational evidences. However, at lower radio frequency bands, the detection of faint photons becomes increasingly challenging for optical telescopes. The limited observational data in this range introduces analytical difficulties, but it also presents significant opportunities for groundbreaking discoveries in new physics beyond the Standard Model (SM).

Notably, the Absolute Radiometer for Cosmology, Astrophysics and Diffuse Emission (ARCADE-2) and other low-frequency surveys have detected an unexpected brightness temperature excess in radio emission at frequency 22MHz-10GHz after subtracting the systematic error and galactic foreground
contamination \citep{haslam1981408,reich1986radio,Roger:1999jy,maeda199945,Fixsen:2009xn}. Fitting over the excess data with a single power-law spectrum gives a frequency dependence of $f^{-2.6}$ \citep{Fixsen:2009xn}. Such a power slope can be approximately realized in the synchrotron emissions from extragalactic sources, although its amplitude accounts for only about one-fifth of the observed excess. Furthermore, the remarkable smoothness of the signal anisotropy indicates that it is likely produced by additional radiation from an unidentified new cosmological source, such as WIMP annihilation \citep{Fornengo:2011cn}, dark photon-photon
oscillations \citep{Caputo:2022keo}, cosmic string decay \citep{Cyr:2023yvj},
primordial black hole emission \citep{Mittal:2021dpe} and so on (see
review \citep{Singal:2017jlh,Singal:2022jaf} and reference therein).

Another significant anomaly in the radio spectrum is the 21cm hydrogen
line absorption signal, detected by the Experiment to Detect the Global EoR Signature (EDGES) in 2018 \citep{Bowman:2018yin}. The absorption trough exhibits an unexpected depth of $-0.5_{-0.5}^{+0.2}$K at a frequency band with the central value 78MHz, which conflicts with the minimum allowed value predicted by the SM at $99\%$ C.L.. 
 	However, re-analyses of the EDGES data raising concerns
 	on the unphysical parameters  for the foreground and   potential presence of unmodelled  systematics  \citep{Hills:2018vyr}. In addition, a follow-up observation in 2021 with Shaped Antenna measurement
		of the background RAdio Spectrum 3 (SARAS-3) announced the nonexistence of such a strong absorption signal, indicating a rejection of the best-fitting
		profile reported by EDGES with $95.3\%$ confidence \cite{Singh:2021mxo}. Recently, in \citep{Sims:2022nwg,Sims:2025kre} EDGES group develops a new Bayesian approach for an unbiased data analysis, and demonstrate the Intrinsic
		model used in \citep{Hills:2018vyr}  is biased. Meanwhile, the on-going third phase of EDGES experiment since 2022 has released the data consistent with the result reported in 2018.	
	Moreover, it is interesting to mention that the very recent observation of JWST UV excess might hint at a deeper 21cm trough than SM \citep{Hassan:2023asd}. 
		Despite the inconclusive results on 21cm signal, in this paper we concentrate on the anomaly data reported from EDGES. 
	 To explain this possible discrepancy, 
Two  main categories have been widely explored: cooling of the
gas temperature through interaction with cold dark matter \citep{Barkana:2018lgd,Berlin:2018sjs,Munoz:2018pzp};
heating the radiation background by additional photon injection from
dark photon \citep{Pospelov:2018kdh}, cosmic string \citep{Brandenberger:2019lfm},
light dark matter, and particularly from ALPs \citep{Moroi:2018vci,Choi:2019jwx}.

Given the possible solutions to both anomalies, a natural question emerges: Is it feasible to simultaneously account for both signals through additional photon injection, and if so, what might be the common source of these photons? This inquiry has only been partially addressed in the literature by examining the influence of radio excess on 21cm physics.
For instance,   the results in Refs. \citep{Feng:2018rje,Mittal:2021egv}
suggest that approximately up to $5\sim8\%$ of the excess data reported by ARCADE-2 and relevant experiments interpolated at 78MHz can produce a significant deep trough without exceeding EDGES observational bounds. These constraints can be significantly relaxed considering the energy transfer between photons and the intergalactic medium through the Lyman-$\alpha$ heating process \citep{Fialkov:2019vnb,Caputo:2020avy}.
Moreover, Ref.  \citep{Acharya:2023ygd} discusses the Bremsstrahlung heating effect caused by soft photons, possibly reconciling two signals. Nonetheless, 
the existing literature does not provide straightforward, unified scenarios to account for both the excess radio spectrum and the deep 21cm absorption. Recent attempts by Refs. \citep{Chianese:2018luo,Dev:2023wel} explore
the decay of relic neutrinos, but the predicted absorption depth fails to match EDGES observations.

\vspace{0.1cm}

In this letter, we propose a unified explanation for the anomalous radio signals of ARCADE-2 and EDGES by considering ALP-photon conversion, attributing the additional photons to ALP dark radiation. 
The resonant
conversion within the ALP mass range $10^{-14}\sim 10^{-12}$eV is characterized by a power-law spectrum with a scaling of $f^{-2}$. This model provides a better fit to the ARCADE-2 data,
more than $5\sigma$ from the null new physics hypothesis, 
suggesting that
the injected photons account for around $2\%$ to $20\%$ of observed radio excess at 78MHz.

\begin{figure*}[t]
\includegraphics[scale=0.6]{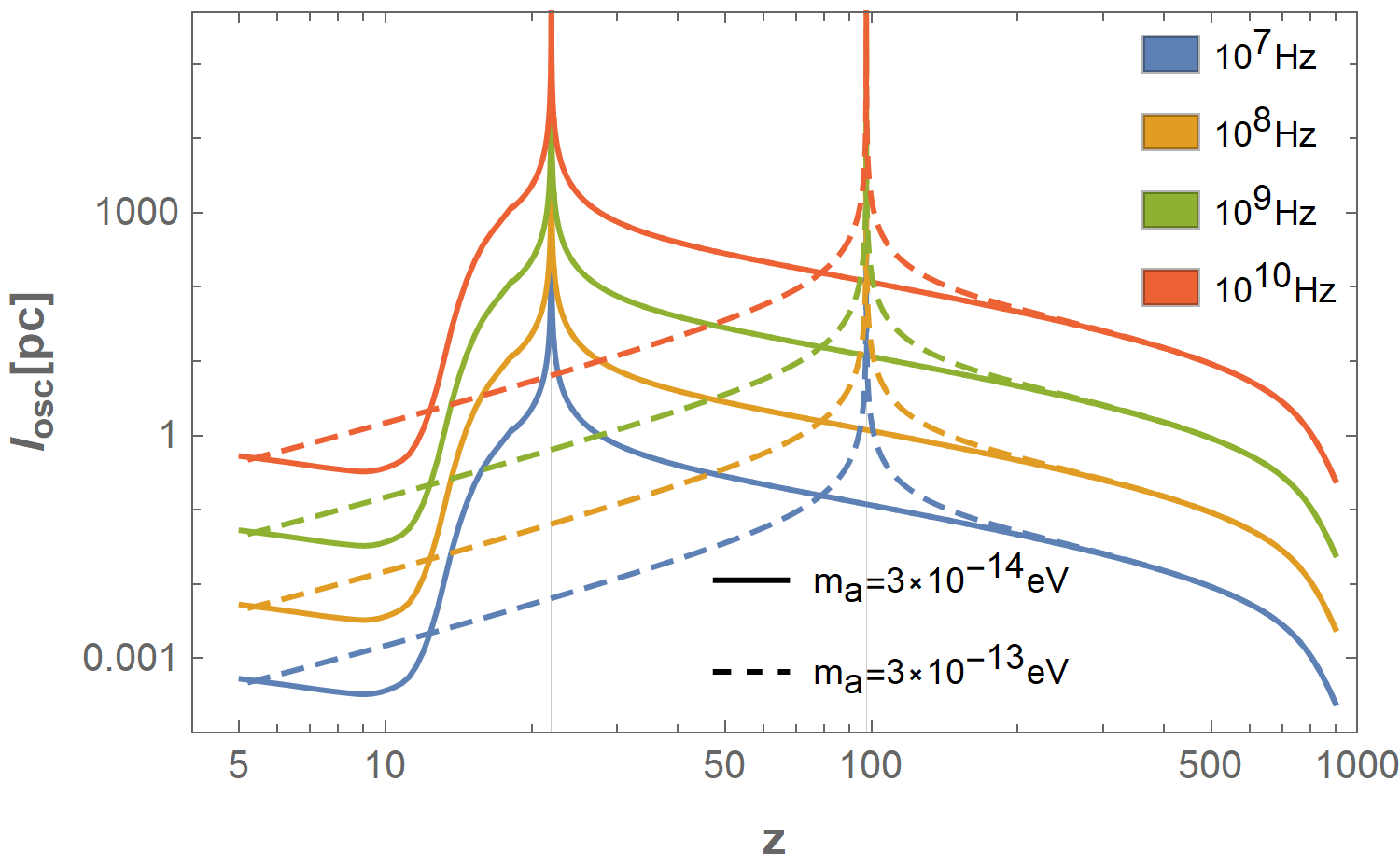}\hspace{0.6cm}\includegraphics[scale=0.63]{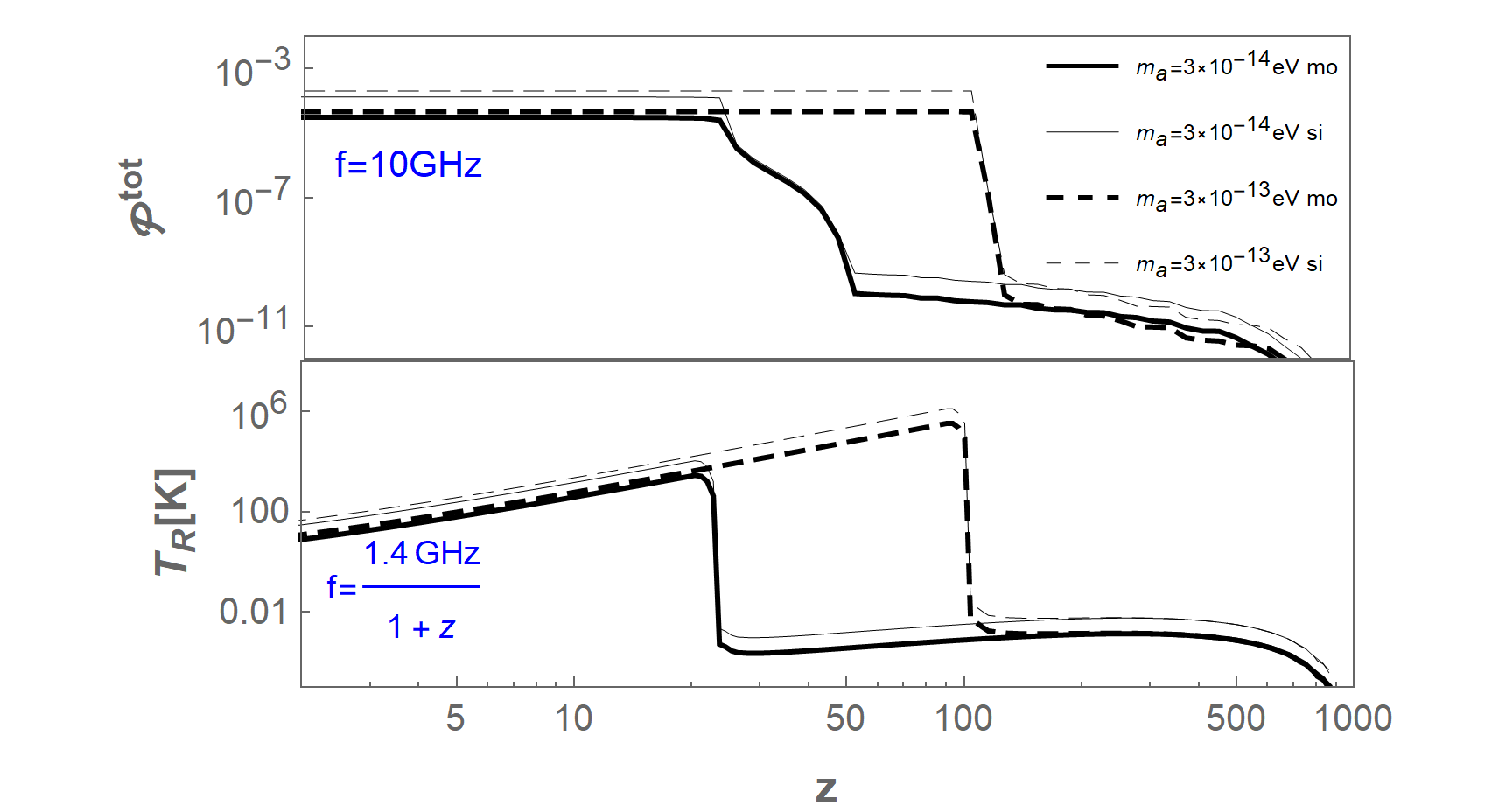}

\caption{\textbf{Left}: the oscillation length profile for two cases with $m_{a}=3\times10^{-14}\textrm{eV}$
and $m_{a}=3\times10^{-13}\textrm{eV}$. The peaks indicate resonances
at the effective mass equality condition $m_{\gamma}=m_{a}$. \textbf{Right}: total probability
of ALP conversion to photon at a fixed frequency $f=10\textrm{GHz}$
(Eq. \ref{eq:pro}) and corresponding brightness temperature
of the radiation from conversion with frequency $f=f_{\textrm{21cm}}/(1+z)$
(Eq. \ref{eq:Tb-z}) . We set $B_{0}=0.1$nGs and $k_{B}^{-1}=0.01$Mpc
for the monochromatic (mo) spectrum of magnetic field and $k_{UV}^{-1}=0.01$Mpc,
$k_{IR}^{-1}=1$Mpc for the truncated scale invariant (si) case.}

\label{fig-losc-P-T}
\end{figure*}

\textbf{\emph{ALP-Photon Mixing.}}
Let us consider the ALP-photon mixing, in expanding Universe, through the interaction term 
\begin{eqnarray}
L_{a\gamma}=-\frac{1}{4}\int d^{4}x\sqrt{-g}g_{a\gamma}F_{\mu\nu}\tilde{F}^{\mu\nu}a\, .
\end{eqnarray}
Here $g$ represents the determinant of flat Friedmann-Lema\^itre-Robertson-Walker (FLRW) metric, $a$ is the ALP
field with mass $m_{a}$, $F_{\mu\nu}$ ($\tilde{F}^{\mu\nu}$) is
the (dual) field strength tensor of electromagnetic potential $A_{\mu}$,
and $g_{a\gamma}$ is the ALP-photon coupling costant. For relativistic
ALP at high frequency $(m_{a}\ll\omega)$, the dynamics of the ALP-photon
mixing along a propagation $l$-direction, perpendicular to $x$-$y$
plane, in the presence of an inhomogeneous magnetic background field
is given by  $\partial_{l}\left(A_{x},A_{y},a\right)^{T}=i\mathcal{K}\left(A_{x},A_{y},a\right)^{T}$ with kernel matrix \citep{Mirizzi:2007hr}
\begin{equation}
 \mathcal{K} = \frac{1}{1+z}\left(\begin{array}{ccc}
		\omega+\Delta_{pl} & 0 & \frac{1}{2}g_{a\gamma}B_{x}(l)\\
		0 & \omega+\Delta_{pl} & \frac{1}{2}g_{a\gamma}B_{y}(l)\\
		\frac{1}{2}g_{a\gamma}B_{x}(l)\qquad & \frac{1}{2}g_{a\gamma}B_{y}(l)\hfill & \omega+\Delta_{a}
	\end{array}\right),\label{eq:axp-eom}
\end{equation}
where $z$ is the redshift of the expanding Universe. In this derivation,
we treat the primordial magnetic field as a perturbation significantly
suppressed by highly isotropy and homogeneity of the Universe, allowing
us to neglect the birefringence and Faraday rotation effect (see Appendix
for more details) \citep{Ejlli:2016asd}. Moreover, the photon propagating in a uniform medium can obtain an effective mass $m_{\gamma}$ defined by the plasma frequency $m_{\gamma}=\omega_{pl}=\sqrt{e^{2}n_{e}/m_{e}}$, where
$e$ is the electron charge, $m_{e}$ the mass and $n_{e}=n_{b0}(1+z)^{3}X_{e}(z)$ the
free electron density with the baryon number density $n_{b0}\simeq0.251\textrm{m}^{-3}$
and ionization fraction $X_{e}(z)$ in intergalactic medium. This
plasma effect is incorporated in the mixing matrix by mass term $\Delta_{pl}=-m_{\gamma}^{2}/2\omega$,
while the ALP mass contribution is $\Delta_{a}=-m_{a}^{2}/2\omega.$ 
In this system, the off-diagonal terms of $\mathcal{K}$ represent the ALP-photon
mixing, enabling possible conversion among ALPs and photons. The
conversion probability is characterized by 
the comoving oscillation length $l_{osc}(z)=2(1+z) |\Delta_{pl}-\Delta_{a}|^{-1}$.
Generally, the schematic relation $\mathcal{P}\propto g_{a\gamma}^{2}B^{2}l_{osc}^{2}$
indicates that a longer oscillation length enhances the conversion
probability. Moreover, a resonance is excited when the effective photon
mass matches the ALP mass, $m_{\gamma}(z_{res})=m_{a}(z_{res})$,
at a certain redshift $z_{res}$. Such a resonance dramatically enhances
the oscillation length, producing a peaked profile as illustrated
in Fig. \ref{fig-losc-P-T} for two benchmark cases: $m_{a}=3\times10^{-14}\textrm{eV}$
and $m_{a}=3\times10^{-13}\textrm{eV}$. In addition, due to the higher ionizing fraction before recombination, the oscillation length is shorter, making ALP-photon conversion negligible. Therefore, our analysis
focuses on ALP-photon oscillations in the post-recombination epoch. 
The other important factor in ALP-mixing is the magnetic field
in the extra-galactic space, which is expected to be characterised by stochastic perturbations.
This can be modeled as a nearly isotropic Gaussian distribution with
the power spectrum $P_{B}(k)$ defined by \citep{Durrer:2013pga}
\begin{eqnarray}
	\left\langle \mathbf{B}_{i}(z,\mathbf{x})\mathbf{B}_{j}\left(z,\mathbf{x}'\right)\right\rangle  & =   \frac{(1+z)^{4}}{(2\pi)^{3}}\int_{k_{IR}}^{k_{UV}}d^{3}ke^{i\mathbf{k}\cdot\left(\mathbf{x}^{\prime}-\mathbf{x}\right)} \nonumber\\
	&  \left(\left(\delta_{ij}-\hat{k}_{i}\hat{k}_{j}\right)P_{B}(k)\right),\label{eq:B-ps}
\end{eqnarray}
where  infrared and ultraviolet cutoffs $k_{IR}$ and $k_{UV}$
appear as extremes of the integral domain delineating the physical limits from
magnetogenesis and field evolution.

In the expanding Universe, the non-commutativity of kernel $\mathcal{K}$ at different
redshifts $[\mathcal{K}(z),\mathcal{K}(z')]\neq0$ does not allow to obtain the analytical
solution of Eq. \ref{eq:axp-eom}. Thus, we use a semi-steady approximation
by discretizing the post-recombination epoch into redshift intervals
$[z_{i},z_{i+1}]$ and  derive the conversion probability
\begin{eqnarray}
	\mathcal{P}(z_{i},z_{i+1})  = & \frac{3\pi}{4(2\pi)^{2}}\frac{(1+z_{c})^{3}g_{a\gamma}^{2}}{H_{c}|\Delta'_{pl}(z_{c})-\Delta'_{a}(z_{c})|}  \nonumber \\
	&\times \int_{k_{IR}}^{k_{UV}}dkk^{2}P_{B}(k)W(t_{1},t_{2};k),\label{eq:pro}
\end{eqnarray}
where $z_{c}=(z_{i}+z_{i+1})/2$, the prime denotes the derivative
with respect to redshift, and $W(t_{1},t_{2};k)$ is specified in Appendix A. Interpolating overall $z_{i}$-sequence yields
a continuous profile of total conversion probability $\mathcal{P}^{tot}(z)$
from recombination to any redshift $z$. In Fig. \ref{fig-losc-P-T},
the conversion probability is shown for two typical magnetic spectra:
a monochromatic spectrum $P_{B}=\pi^{2}B_{0}^{2}\delta\left(k-k_{B}\right)/k_{B}^{2}$
and a truncated scale-invariant spectrum $P_{B}=\pi^{2}B_{0}^{2}/k^{3}$
with cutoffs $k_{IR}\leq k\leq k_{UV}$, where $B_0$ is the magnetic field at present day. Notably, the probability increases
sharply when mass-equal resonance occurs at $z_{res}\simeq20$ for
$m_{a}=3\times10^{-14}$eV and $z_{res}\simeq100$ for $m_{a}=3\times10^{-13}$eV,
correlated with oscillation length peaks in 
Fig. \ref{fig-losc-P-T}. Moreover, the amplification of probability
in $m_{a}=3\times10^{-14}$ eV case begins as early as $z\simeq50$
preceding the mass-equal resonance at $z_{res}\simeq20$. This is caused
by the stochastic resonance excited when $\lambda_{B}\lesssim l_{osc}$
for the monochromatic spectrum and $k_{UV}^{-1}\lesssim l_{osc}$
for the truncated invariant spectrum  \citep{Addazi:2024kbq}.

\textbf{\emph{Radio Excess and 21cm Trough.}}
We consider relativistic ALPs produced in the early Universe and analyze their impact on the radio spectrum via ALP-photon oscillation throughout
the post-recombination era. Since the relevant frequencies lie in the Rayleigh-Jeans regime ($\omega\lesssim$THz), we express the
radiation as brightness temperature for direct comparison with observations.
Assuming a frequency-independent ALP abundance, the cumulative extra
radiation from ALPs at comoving frequency $\omega_{0}$ till the present
day is given by 
\begin{equation}
T_{b0}^{AP}(\omega_{0})  =  \frac{\pi^{4}\gamma}{15}\frac{T_{0}^{4}}{\omega_{0}^{3}}\mathcal{P}^{tot}(\omega_{0},z=0),\label{eq:Tb-ARCADE}
\end{equation}
where $T_{0}=2.725$K is the CMB temperature
and $\gamma=\boldsymbol{\Omega}_{a}/\boldsymbol{\Omega}_{\gamma}$
represents the present-day ALP-to-photon energy density ratio. The ALP energy density $\boldsymbol{\Omega}_{a}$ is constrained by the
extra effective relativistic species $\triangle N_{\textrm{eff}}$
with $\boldsymbol{\Omega}_{a}/(0.23\boldsymbol{\Omega}_{\gamma})\lesssim\triangle N_{\textrm{eff}}$.
The \textit{Planck} constraint $\triangle N_{\textrm{eff}}\lesssim0.33$
gives rise to the bound $\gamma\lesssim0.06$ \citep{Planck:2015zrl}.
We set $\gamma=0.06$ in this work. In addition
to the ALP mass range $10^{-14}$-$10^{-12}$eV, we choose coupling $g_{a\gamma}=5\times10^{-13}{\rm GeV}^{-1}$ which is slightly beyond  the  region excluded 
by \textit{Chandra} observation of the quasar \citep{Reynes:2021bpe}, and can be potentially probed
by DANCE \citep{Obata:2018vvr} and Twisted Anyon Cavity \citep{Bourhill:2022alm}.

\begin{figure*}[t]
\includegraphics[scale=0.63]{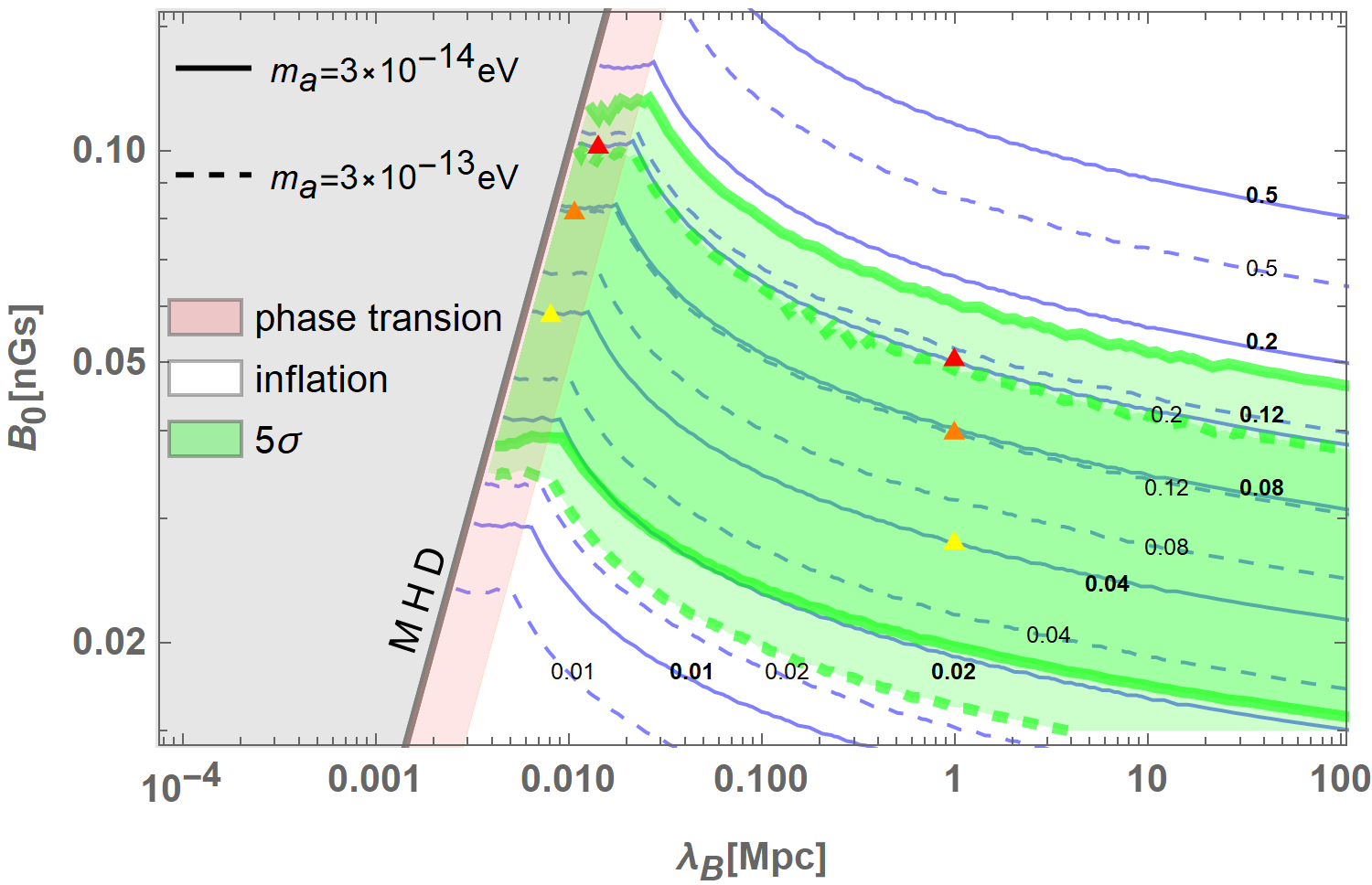} \, \,\,\,\,\hspace{0.5cm} \includegraphics[scale=0.63]{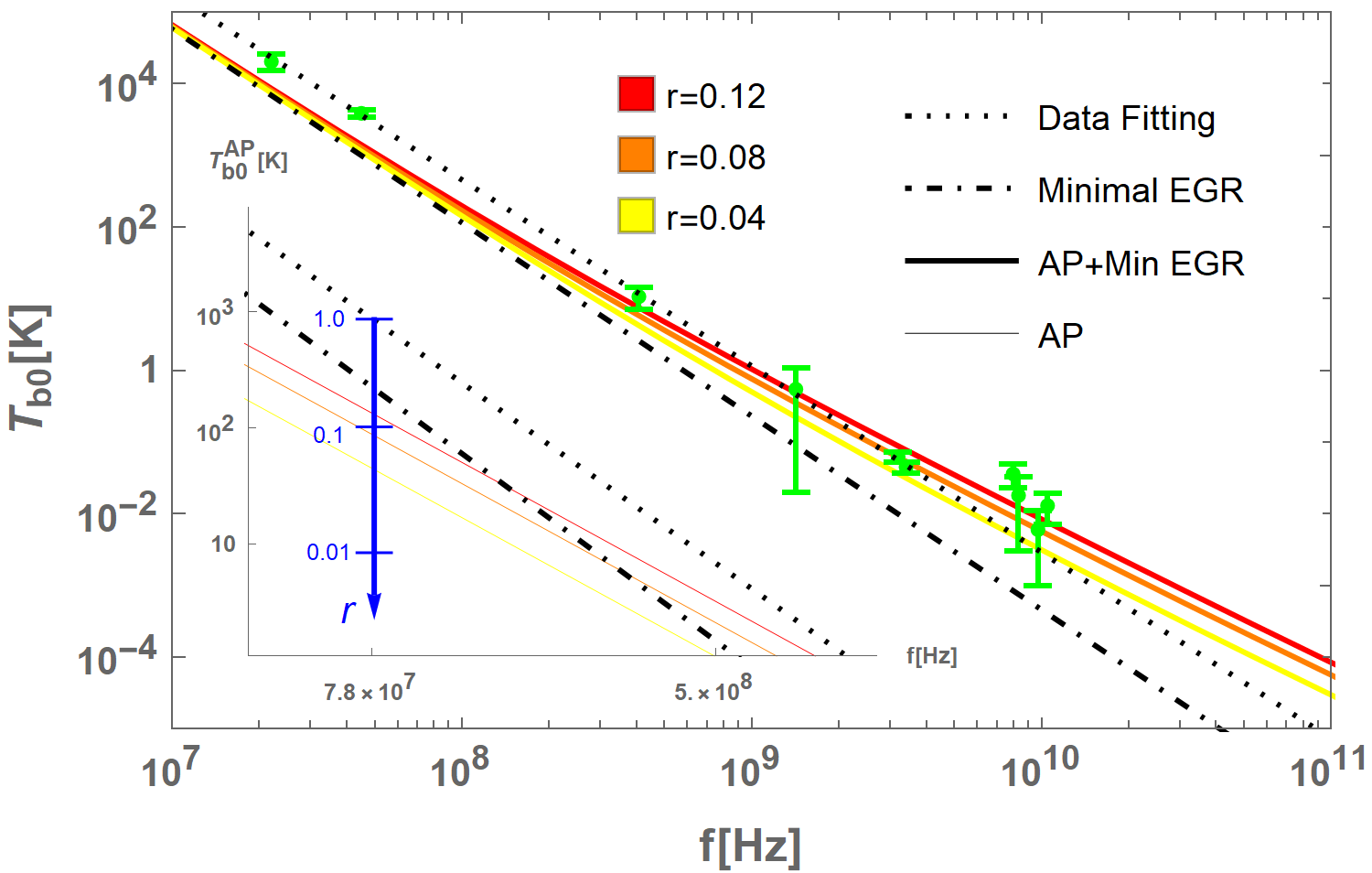}

\caption{\textbf{Left}: Overlap of the fractional contribution from the ALP-photon conversion to the observed excess radio at $78$MHz (blue contours) and the $5\sigma$ improvement over the model incorporating only astrophysical synchrotron source
(green shade). The gray area represents the exclusion
of the primordial magnetic field relics based on MHD analysis. The pink and white regions correspond to the relic magnetic field with peaked and scale-invariant spectrum, respectively. \textbf{Right}: radio
spectrum benchmarks for ALP-photon conversion with $m_{a}=3\times10^{-14}$eV
in addition to the minimal extra-galactic radiation, corresponding to the triangles in the upper-left panel. Green error bars are radio excess
observed by ARCADE-2 ($3$-$90$GHz) and other low-frequency measurements.
}

\label{fig-experiments}
\end{figure*}

We model out the cosmic magnetic field with a two-scaling spectrum as \citep{Brandenburg:2018ptt}
\begin{eqnarray}
    P_{B}\propto(k/k_{0})^{\alpha}\,\,{\rm with}\,\,\begin{cases}
     \alpha=n_L\,\,{\rm for}\,\, k_{IR}<k\leq k_{0}, \\ \alpha=n_S\,\,{\rm for}\,\, k_{0}<k\leq k_{D},
 \end{cases}\label{ps}
\end{eqnarray}
where $n_{L}$ and $n_{S}$ are spectral indices for large and small scales, respectively. The field is processed by magnetohydrodynamics (MHD) at small scales, resulting in turbulence with a universal Kolmogorov scaling of $n_{S}=-11/3$ \citep{kolmogorov1991local}.
The damping scale $k_{D}$, acting as the ultraviolet cutoff,  is determined by the plasma viscosity during
recombination, approximately $k_{D}\simeq\mathcal{O}(100)\left(10^{-9}{\rm Gs}/B_{0}\right){\rm Mpc^{-1}}$
\citep{Kahniashvili_2010}. On large scales, the magnetic field retains
its initial configuration, carrying information about magnetogenesis.
Two types of primordial magnetogenesis are widely studied in the literature:
from phase transition with $n_{L}=2$ and the inflation with $n_{L}=-3$
\citep{Durrer:2013pga}. For the latter with a truncated scale-invariant spectrum, the infrared cutoff $k_{IR}$  marks the beginning of magnetogenesis during inflation.  Properly taking into account the MHD evolution
excludes certain regions in the $B_{0}$-$\lambda_{B}$ (gray shade
in Fig. \ref{fig-experiments}). The relic field from phase transition
with a peaked spectrum survives in a narrow band near the MHD line
(region illustrated in pink shade), while relic field from inflation
with a  scale-invariant spectrum could fill the broad
region right to the MHD line.

We employ Eqs. \ref{eq:pro} and \ref{eq:Tb-ARCADE} to simulate the
total radiation from the ALP conversion till today. We run
over the parameter space in allowed $B_{0}$-$\lambda_{B}$ region
and find that brightness temperature of the radiation scales simply
as $T_{b0}^{AP}\propto\omega_{0}^{-2}$. For a comparison with 21cm study (as we will see later), we define $r=T_{b0}^{AP}/T_{b0}^{obs}|_{f_{78}}$
as the fractional contribution from the ALP-photon conversion to the
observed radio excess temperature $T_{b0}^{obs}$ extrapolated at
$f_{78}=78$MHz. In  Fig. \ref{fig-experiments},
we plot $r$-contours over the $B_{0}$-$\lambda_{B}$ space for both
peaked and truncated scale-invariant spectra with two ALP mass cases
$m_{a}=3\times10^{-14}$eV (solid) and $3\times10^{-13}$eV (dashed).
With the informations on the predicted extra radiation injected from ALP, we are
ready to compare it with the observational radio excess in the CMB Rayleigh-Jeans
tail. We perform the $\chi^{2}$-analysis across 14 experimental data
from low-frequency survey (22, 45, 408, 1420MHz) \citep{haslam1981408,reich1986radio,Roger:1999jy,maeda199945},
ARCADE-2 (3.2, 3.41, 7.98, 8.33, 9.72, 10.49, 29.5, 31, 90GHz)\citep{Fixsen:2009xn}
and the CMB temperature measured by FIRAS (250GHz) \citep{fixsen2002spectral}.
For $m_{a}=3\times10^{-14}$eV case, we run over the allowed $B_{0}$-$\lambda_{B}$
parameter region and obtain the best fit with a minimum $\chi_{min}^{2}=49$
($12$ d.o.f) by counting the contribution from ALP-photon conversion
plus the irreducible extra-galactic emission. The inclusion of the
extra radiation converted from ALP significantly improves over $\chi^{2}=114$
($12$ d.o.f) from the extra-galactic contribution alone. In  Fig. \ref{fig-experiments}, we show 
	a region achieving $5\sigma$ improvement over the null-hypothesis model incorporating only extra-galactic source, which can be considered  as the
	viable parameter space providing a potential explanation for
	the radio excess. 
To have a straightforward demonstration, we show
several benchmarks with values at the triangle points in the upper-left
panel, among whom the $r=0.08$ case
corresponds to the best fit. For comparison, the radio spectrum from
a purely extragalactic synchrotron source is shown in a dot-dashed line
with the label ``Minimal EGR''. Indeed, benchmarks within our ALP model
can pass through most of excess data in the $0.4$-$10$GHz range, explaining observed
excess in the relevant band, particularly in the ARCADE-2 band. Moreover,
the $5\sigma$- viable space (green shade) almost overlaps with the
region bounded by $r\simeq 0.02$ and $r\simeq0.2$. This coincidence can be
easily seen because, in the viable space, the ALP-induced radiation
dominates over the emission from the extra-galactic sources at temperature
$f\gtrsim0.1$ GHz. Results for $m_{a}=3\times10^{-13}$ eV case are
similar.

\begin{figure*}[t]
	\includegraphics[scale=0.7]{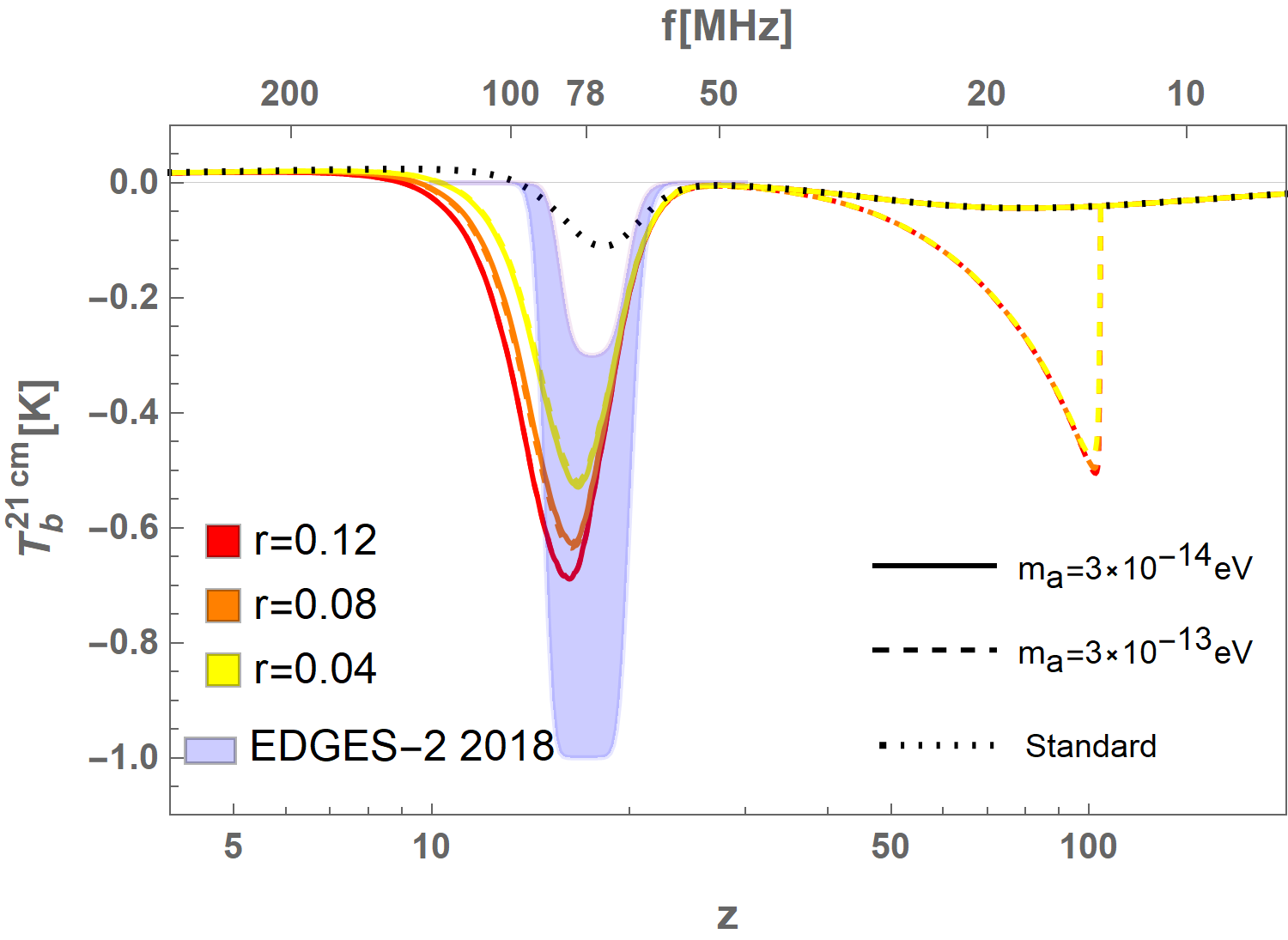}
	\includegraphics[scale=0.7]{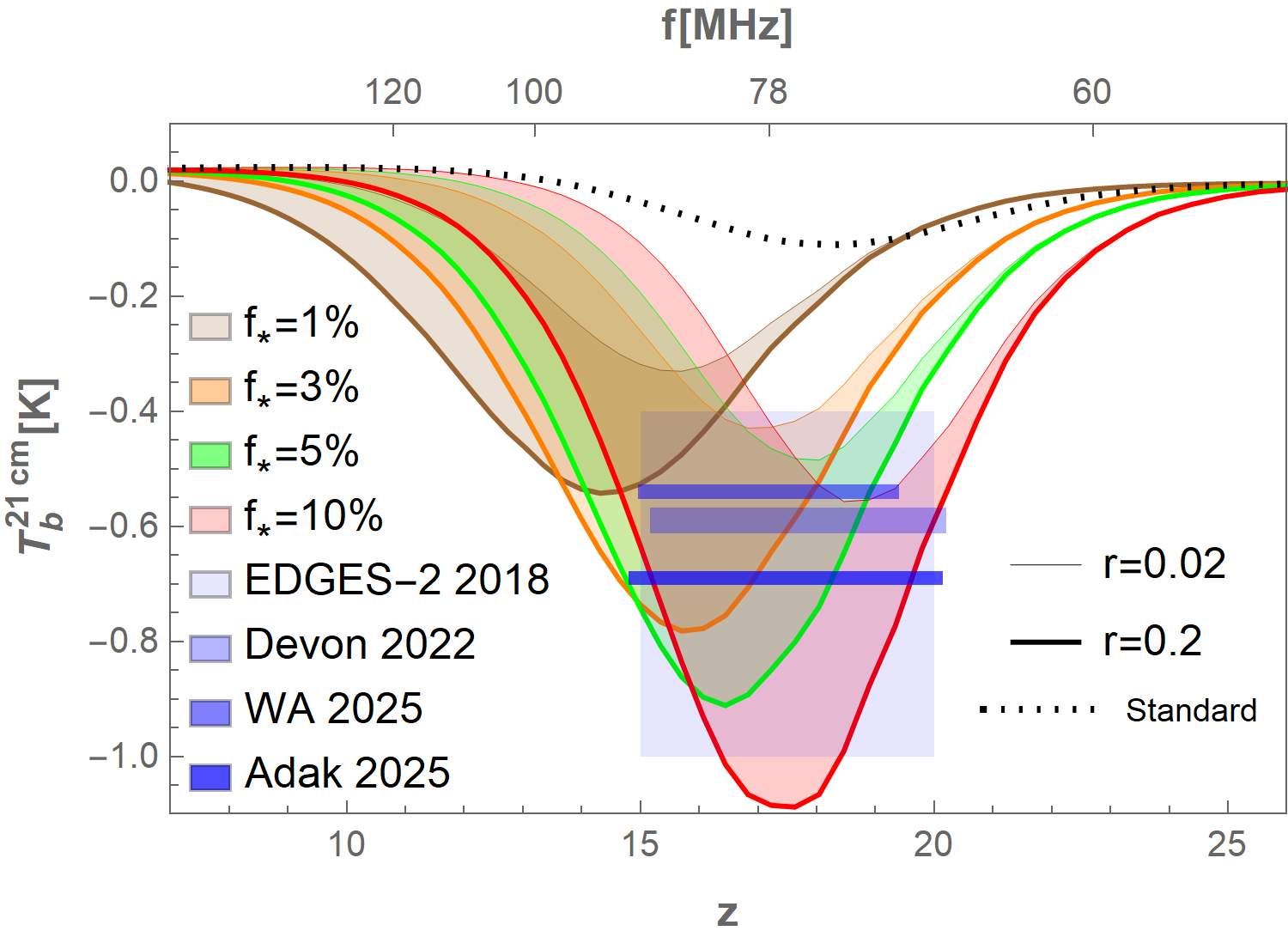}
	
	\caption{\textbf{Left:} Benchmark model with star formation efficiency $f_*=3\%$  for 21cm global signal under the modified radiation background, compared with EDGES-2 observation in 2018. Three typical values of $r$ correspond to benchmark of ARCADE-2 in Fig. \ref{fig-experiments}. \textbf{Right:} profiles of 21cm lines generated with different $f_*$ and in range  $0.02<r<0.2$ referred from the  $5\sigma$ region bound of ARCADE-2 in Fig. \ref{fig-experiments}. The blue shades denote the data reported in  EDGES-2,3 observations (see main text for details).  
	}
	
	\label{fig-21cm}
\end{figure*}

Besides the direct influence on the radio excess in the present day, the extra photons from ALP conversion can also affect the background temperature throughout the post-recombination era and hence impact on 21cm physics. At redshift $z$, we count the extra photons
at physical frequency $f_{\rm 21cm}=1.4$ GHz arising from the ALP conversion
as
\begin{eqnarray}
T_{b}^{AP}(z) & = & \frac{\pi^{4}\gamma}{15}\left(\frac{T_{0}}{\widetilde{\omega}}\right)^{3}\mathcal{P}^{tot}(\widetilde{\omega},z)T_{0}(1+z),\label{eq:Tb-z}
\end{eqnarray}
with $\widetilde{\omega}=2\pi f_{\rm 21cm}/(1+z)$. Due to the narrow resonance at mass-equal redshift $z_{res}$, the
brightness temperature rapidly increases by several orders of the
magnitude in a short period, as shown in  Fig. \ref{fig-losc-P-T}.
After resonance, the scaling of   today's temperature $T_{b0}^{AP}\propto f^{-2}\propto(1+z)^{2}$ indicates the scaling  $T_{b}^{AP}(z)\propto(1+z)^{3}$  at certain redshift (see Fig. \ref{fig-losc-P-T} ). Additionally, the temperature amplitude   can be normalized at 78MHz by the parameter $r$.
Thus in practice, we can simply parameterize the full background temperature
inclusion of CMB radiation as $T_{R}(z)=T_{0}(1+z)+rC\left(1+z\right)^{3}\Theta\left(z_{res}-z\right),$where
$C=\left(f_{\rm 21cm}/f_{78}\right)^{-2}T_{b0}^{obs}|_{f_{78}}=2.698$K
and $\Theta(z)$ is the Heaviside function. In this setup, only two
free parameters are relevant for 21cm analysis, namely $z_{res}$
determined by $m_{a}$ (see Fig. \ref{fig-losc-P-T}) and
$r$ with distribution shown in $B_{0}$-$\lambda_{B}$ plane (see Fig. \ref{fig-experiments}). Since the viable space to explain
the radio excess roughly spans $r\simeq 0.02$ to $0.2$ contours, we
pick three representative choices $r=0.04,0.08$ and $0.12$ to parameterize
the modified background temperature. We simulate the 21cm global signal
as in Ref. \citep{Caputo:2020avy} and compare with EDGES data. The
same model for Lyman-$\alpha$ and X-ray heating as in
Ref. \citep{Caputo:2020avy} and reference therein is employed, where energy transfer
between modified radiation background and intergalactic medium is
taken into account \citep{Fialkov:2019vnb}. 
We consider several reasonable values of star formation efficiency $f_*$
at $1\%$, $3\%$, $5\%$ and $10\%$. Additionally, the heating from Lyman-$\alpha$ and X-ray
flux are respectively controlled by the normalized emissivity factor
$f_{\alpha}$ and $f_{X}$, which are expected to be around the order of unity in general stellar dynamics \citep{Pritchard:2011xb}. We set the fiducial values $f_{\alpha}=f_{X}=1$ and present numerical results in Fig. \ref{fig-21cm} (see Appendix C for non-fiducial cases). To demonstrate the possibility of fitting anomalous data of ARCADE-2 and EDGES-2 (EDGES in 2018 \citep{Bowman:2018yin}) simultaneously,  we choose $f_*=3\%$ and  $r=0.04,0.08$ and $0.12$ corresponding to benchmarks in Fig.\ref{fig-experiments}. As shown in the left panel, the photon injection from ALP
conversion in benchmarks of ARCADE-2  can indeed produce a sufficiently
deep absorption trough that aligns with the EDGES-2  measurement. 
Intriguingly,
the benchmark with $m_{a}=3\times10^{-13}$eV exhibits an additional
absorption trough deeper than the Standard Model prediction at a lower
frequency $f\sim15$MHz. This is caused by the ALP-photon oscillation
resonance at the higher redshift $z\sim100$. In contrast, this feature
is absent in the benchmark with $m_{a}=3\times10^{-14}$eV  because of the resonant photon injection occurring at lower redshift $z \sim 20$
(see Fig. \ref{fig-losc-P-T}). 

Regarding the observation on 21cm global signal, EDGES group has carried on their experiment to the third phase, dubbed EDGES-3, consisting of state-of-art  data analysis technology and multiple-sites  detectors deployed further to  Devon Island  and Adak Island besides the original zone in western Australia (WA).  According to the lab memo, the updated data supports the deep trough consistent with EDGES-2 measurement in 2018 \footnote{\url{https://www.haystack.mit.edu/haystack-memo-series/edges-memos/} }. In particular, some subtraction tests within only 4 terms to model the foreground can be a solid evidence that the observed deep absorption is a true physical signal rather than systematic bias \footnote{\url{https://www.haystack.mit.edu/wp-content/uploads/2025/01/memo_EDGES_466.pdf}}. We adopt the EDGES-2 result, together with the recent preliminary results including one data from Devon Island in 2022 and two from WA and Adak Island in 2025 \footnote{\url{https://www.haystack.mit.edu/wp-content/uploads/2025/05/edgesmemo_481.pdf}}, and show them in blue shade regions in right panel of Fig. \ref{fig-21cm}. We also need to superimpose 21cm signals predicted in our ALP-photon conversion model. To this end, we first notice that the profile of the 21cm trough around $z\sim 17$  in our model is mainly governed by the star formation efficiency $f_*$ and fraction parameter $r$, where the latter quantifies the excess radiation from ALP conversion. In the right panel, we plot 21cm line with different combinations of $f_*=1,3,5,10\%$ and $r$ in range $0.02<r<0.2$, which is the bound of $5\sigma$ improvement on ARCADE-2 excess fitting (see Fig.\ref{fig-experiments}). We find an extensive overlap area between the adopted data and model prediction, indicating that the proper amount of photon converted from ALP could potentially provide a simultaneous explanation on ARCADE-2 and EDGES-2,3 anomalies.  This conclusion should be considered provisional for reference only since data from EDGES-2,3 are still inconclusive. 

Nevertheless,  provided the confirmed 21cm observational data in future, it can inversely constrain on $r$ (assume $f_*$ is estimated from other astrophysical observation), and hence on the existence/properties of ALP as well as the primordial magnetic field.  Generally speaking, shallow 21cm trough with $T_b\lesssim 300$mK (including the SM case) implies $r\lesssim0.02$,  indicating a failure of our model to explain ARCADE-2 excess. For deeper trough that can be either inside or outside the EDGES-2,3 region, the ALP-photon conversion can be a possible solution to both ARCADE-2 excess and non-standard 21cm signal. As shown in  Fig. 6, given  the value of   $r$ from the true 21cm data, one can infer the mass and  ALP-photon coupling of  ALP,  and the strength and correlation length of primordial magnetic field.




\textbf{\emph{Conclusion.}}
In this letter, we explored the mixing of axion-like particles (ALPs) with photons in the presence of a stochastic magnetic field during the post-recombination epoch, specifically focusing on resonance effects for ALP masses in the range of $10^{-14}\sim 10^{-12}$eV. Our analysis targeted two anomalous phenomena within the Rayleigh-Jeans tail of the CMB: the radio excess detected by ARCADE-2 and other low-frequency surveys, as well as the pending-confirmation unexpected 21cm absorption trough observed by EDGES-2,3. 


We determined feasible parameter regions for the magnetic field that allow for a simultaneous explanation of both anomalies, with several benchmarks presented in Figs. \ref{fig-experiments}-\ref{fig-21cm}. 
From another perspective, the joint analysis of the CMB excess and 21cm absorption provides a novel probe into the properties of the dark sector and primordial magnetogenesis.


\textbf{\emph{Acknowledgment.}} We would like to thank Luca Visinelli and Theodoros Papanikolaou for useful discussions and suggestions.
AA work is supported by the National Science Foundation of China (NSFC) through the grant No. 12350410358; the Talent Scientific Research Program of College of Physics, Sichuan University, Grant No. 1082204112427; the Fostering Program in Disciplines Possessing Novel Features for Natural Science of Sichuan University, Grant No.2020SCUNL209 and 1000 Talent program of Sichuan province 2021.
 SC, GL, and QG    acknowledge the support of Istituto Nazionale di Fisica Nucleare, Sez.\ di Napoli and Gruppo Collegato di Salerno,  Iniziative Specifiche QGSKY and MoonLight-2, Italy.
 RS acknowledges the support of the research project TAsP (Theoretical Astroparticle Physics) funded by the Istituto Nazionale di Fisica Nucleare (INFN).  This article is based upon work from COST Action CA21136 Addressing observational tensions in cosmology with systematic and fundamental physics (CosmoVerse) supported by COST (European Cooperation in Science and Technology). This article is based upon work from the COST Actions “COSMIC WISPers” (CA21106).
 
\begin{widetext} 
\section*{Appendix A}

In this appendix we derive the main formula of our paper. We consider
an ALP-photon mixing system propagating along the $l$-direction in
an perturbative magnetic background field $\boldsymbol{B}(l)=(B_{x},B_{y},B_{z})$.
For highly relativistic axions $m_{a}\ll\omega$,
	the short-wavelength approximation can be applied and the ALP-photon
mixing in the expanding universe reduces to a linearized system \citep{Mirizzi:2007hr}
\begin{eqnarray}
	\left(\omega-i(1+z)\frac{\textrm{d}}{\textrm{d}l}+M\right)\left(\begin{array}{c}
		A_{x}\\
		A_{y}\\
		a
	\end{array}\right) & = & 0,\qquad M=\left(\begin{array}{ccc}
		\Delta_{xx} & \Delta_{xy} & \frac{1}{2}g_{a\gamma}B_{x}\\
		\Delta_{xy} & \Delta_{yy} & \frac{1}{2}g_{a\gamma}B_{y}\\
		\frac{1}{2}g_{a\gamma}B_{x}\qquad & \frac{1}{2}g_{a\gamma}B_{y}\qquad & \Delta_{a}
	\end{array}\right)\label{eq:Apen-eom}
\end{eqnarray}
with the scale factor $a$ and the spatial coordinate $l$ in the
comoving framework. Here the mixing matrix element reads
\begin{eqnarray}
	\Delta_{xx} & = & \Delta_{pl}+\Delta_{CM},\quad\Delta_{yy}=\Delta_{pl}+\Delta_{CM},\quad\Delta_{xy}=\Delta_{yx}=\Delta_{R},\quad\Delta_{a}=-m_{a}^{2}/2\omega,\quad\Delta_{pl}=-\omega_{pl}^{2}/2\omega
\end{eqnarray}
where $m_{a}$ is the ALP mass and $g_{a\gamma}$ the ALP-photon coupling
strength. The plasma effect arsing from photon refraction in the medium
is characterized by the plasma frequency $\omega_{pl}=\sqrt{e^{2}n_{e}/m_{e}}$
with free electron's charge $e$, mass $m_{e}$ and the number density
$n_{e}$. In addition, the term $\Delta_{CM}$ 
describe the CottonMouton effect, which originate from QED corrections
for photon in a transverse magnetic field. As shown in Ref. \citep{Ejlli:2016asd},
this effect is quadratic in magnetic strength and hence negligible
in our analysis since we focus on the linear order of perturbative
$B$ field. The Faraday rotation term $\Delta_{R}$ couples the$A_{x}$
and $A_{y}$ modes, which is relevant for analyzing polarized photon
sources but irrelevant to this study. Thus, Eq. \ref{eq:Apen-eom}
simplifies as
\begin{eqnarray}
	\partial_{l}\left(\begin{array}{c}
		A_{x}(\omega,l)\\
		A_{y}(\omega,l)\\
		a(\omega,l)
	\end{array}\right) & = & i\mathcal{K}\left(\begin{array}{c}
		A_{x}(\omega,l)\\
		A_{y}(\omega,l)\\
		a(\omega,l)
	\end{array}\right),\,\, \,\, \hfill \mathcal{K}=\frac{1}{1+z}\left(\begin{array}{ccc}
		\omega+\Delta_{pl} & 0 & \frac{1}{2}g_{a\gamma}B_{x}(l)\\
		0 & \omega+\Delta_{pl} & \frac{1}{2}g_{a\gamma}B_{y}(l)\\
		\frac{1}{2}g_{a\gamma}B_{x}(l)\qquad & \frac{1}{2}g_{a\gamma}B_{y}(l)\hfill & \omega+\Delta_{a}
	\end{array}\right).\label{eq:Apen-eom-sim}
\end{eqnarray}
We apply the perturbative approach by decomposing matrix $\mathcal{K}$ into
\begin{eqnarray}
	\mathcal{K} & = & \mathcal{K}_{0}+\delta \mathcal{K}=\frac{1}{1+z} \left(\begin{array}{ccc}
		\omega+\Delta_{pl} & 0 & 0\\
		0 & \omega+\Delta_{pl} & 0\\
		0 & 0 & \omega+\Delta_{a}
	\end{array}\right)+\frac{1}{1+z} \left(\begin{array}{ccc}
		0 & 0 & \frac{1}{2}g_{a\gamma}B_{x}(l)\\
		0 & 0 & \frac{1}{2}g_{a\gamma}B_{y}(l)\\
		\frac{1}{2}g_{a\gamma}B_{x}(l) & \frac{1}{2}g_{a\gamma}B_{y}(l) & 0
	\end{array}\right).
\end{eqnarray}
Physically speaking, the evolution of ALP and photon free of external
field is described by principle part $\mathcal{K}_{0}$, and their mixing through
the magnetic background is encoded in perturbation matrix $\delta \mathcal{K}$.
Note that $\mathcal{K}_{0}$ implicitly depends on $l$ from relation $dl=dz/H$
in expanding universe. We arrange Eq. \ref{eq:Apen-eom-sim} into
\begin{eqnarray}
	\partial_{l}\left(e^{-i\int_{l_{0}}^{l}dl_{1}\mathcal{K}_{0}\left(l_{1}\right)}\mathcal{U}\left(l,l_{0}\right)\right) & = & ie^{-i\int_{l_{0}}^{l}dl_{1}\mathcal{K}_{0}\left(l_{1}\right)}\delta \mathcal{K}(l)\mathcal{U}\left(l,l_{0}\right)\label{eq:Apen-U-eom}
\end{eqnarray}
by introducing a conversion matrix $\mathcal{U}$ defined by $\left(A_{x}(l),A_{y}(l),a(l)\right)^{T}=\mathcal{U}\left(l,l_{0}\right)\left(A_{x}(l_{0}),A_{y}(l_{0}),a(l_{0})\right)^{T}$.
One can iteratively solve Eq. \ref{eq:Apen-U-eom} up to the first
order
\begin{eqnarray}
	\mathcal{U}\left(l,l_{0}\right) & = & e^{i\int_{l_{0}}^{l}dl_{1}\mathcal{K}_{0}(l_{1})}+ie^{i\int_{l_{0}}^{l}dl''\mathcal{K}_{0}(l'')}\int_{l_{0}}^{l}dl^{\prime}e^{-i\int_{l_{0}}^{l^{\prime}}d\ell^{\prime\prime}\mathcal{K}_{0}(l'')}\delta \mathcal{K}\left(l^{\prime}\right)e^{i\int_{l_{0}}^{l^{\prime}}dl^{\prime\prime}\mathcal{K}_{0}(l'')}+\mathcal{O}\left(\left(\delta \mathcal{K}\right)^{2}\right).\label{eq:Apen-U-sol}
\end{eqnarray}
Each component of $\mathcal{U}\left(l,l_{0}\right)$ reflects the
mode transitions btween ALP and photon states. For instance, photon
generation at distance $l$ can be read from the conversion matrix
with $A_{x}(l)=\mathcal{U}_{13}a(l_{0})$ and $A_{y}(l)=\mathcal{U}_{23}a(l_{0})$,
leading to the conversion probability 
\begin{eqnarray}
	\mathcal{P}(l) & = & \frac{|A_{x}(l)|^{2}+|A_{y}(l)|^{2}}{\left|a(l_{0})\right|^{2}}=|\mathcal{U}_{13}(l,l_{0})|^{2}+|\mathcal{U}_{23}(l,l_{0})|^{2}.\label{eq:Apen-P-exp}
\end{eqnarray}
From Eq. \ref{eq:Apen-U-sol} one can obtain
\begin{eqnarray}
	\left|\mathcal{U}_{13}\left(l,l_{0}\right)\right|^{2} & = & \frac{1}{4}g_{a\gamma}^{2}\int_{l_{0}}^{l}dl_{1}\int_{l_{0}}^{l}dl_{2}\left(\frac{1}{1+z}B_{x}(l_{1})\right)\left(\frac{1}{1+z}B_{x}(l_{2})\right)e^{i\int_{l_{1}}^{l_{2}}dl_{3}\frac{1}{1+z}(\Delta_{xx}(l_{3})-\Delta_{a}(l_{3}))},\nonumber \\
	\left|\mathcal{U}_{23}\left(l,l_{0}\right)\right|^{2} & = & \frac{1}{4}g_{a\gamma}^{2}\int_{l_{0}}^{l}dl_{1}\int_{l_{0}}^{l}dl_{2}\left(\frac{1}{1+z}B_{y}(l_{1})\right)\left(\frac{1}{1+z}B_{y}(l_{2})\right)e^{i\int_{l_{1}}^{l_{2}}dl_{3}\frac{1}{1+z}(\Delta_{xx}(l_{3})-\Delta_{a}(l_{3}))}.\label{eq:Apen-U-comp}
\end{eqnarray}
Regarding the magnetic field, we assume it to be a statistically isotropic
Gaussian distributed random field with correlation function defined
by
\begin{eqnarray}
	\left\langle \mathbf{B}_{i}(\mathbf{x_{1}})\mathbf{B}_{j}\left(\mathbf{x}_{2}\right)\right\rangle =\frac{1}{(2\pi)^{3}}\int_{k_{IR}}^{k_{UV}}d^{3}ke^{i\mathbf{k}\cdot\left(\mathbf{x}_{1}-\mathbf{x_{2}}\right)}\left[\left(\delta_{ij}-\hat{\mathbf{k}}_{i}\hat{\mathbf{k}}_{j}\right)P_{B}(k)-i\epsilon_{ijm}\hat{\mathbf{k}}_{m}P_{aB}(k)\right],
\end{eqnarray}
where $\hat{\mathbf{k}}=\mathbf{k}/k$ and $\epsilon_{ijm}$ is the
antisymmetric symbol. The $P_{B}(k)$ and $P_{aB}(k)$ are respectively
the symmetric and antisymmetric components of the spectrum, whereas
the latter is absent from the expression 
\begin{eqnarray}
	\left\langle \mathbf{B}_{x}(l_{1})\mathbf{B}_{x}\left(l_{2}\right)+\mathbf{B}_{y}(l_{1})\mathbf{B}_{y}\left(l_{2}\right)\right\rangle =\frac{1}{(2\pi)^{3}}\int d^{3}ke^{ik\cos\theta\left(l_{1}-l_{2}\right)}\left(1+\cos^{2}\theta\right)P_{B}(k).\label{eq:Apen-BxBy}
\end{eqnarray}
Here $\theta$ denotes the angle between propagation direction $l$
and the wave-vector $\boldsymbol{k}$ of the magnetic field. Combining
Eqs. \ref{eq:Apen-P-exp}, \ref{eq:Apen-U-comp} and \ref{eq:Apen-BxBy},
we obtain
\begin{eqnarray}
	\mathcal{P} & = & \frac{3}{8(2\pi)^{3}}\left(1+z\right)^{2}g_{a\gamma}^{2}\int d^{3}ke^{ik\cos\theta\left(l_{1}-l_{2}\right)}P_{B}(k)\int_{l_{0}}^{l}dl_{1}\int_{l_{0}}^{l}dl_{2}\exp\left(i\int_{l_{1}}^{l_{2}}dl_{3}\frac{1}{1+z}(\Delta_{pl}(l_{3})-\Delta_{a}(l_{3}))\right)\label{eq:Apen-P}
\end{eqnarray}
with approximate $1+\cos^{2}\theta\simeq3/2$ to simplify the calculation.
In fact, we have checked that inclusion of exact term $1+\cos^{2}\theta$
yields a minor correction ( $\sim1.1$ correction factor in overall
amplitude) to this approximation.
To perform the integral, we use a semi-steady approximation by discritizing
the universe expansion into a decreasing sequence of redshifts $[z_{1},z_{2},...z_{N}]$
with $z_{1}\simeq1100$ marking the end of Recombination. A proper
discretization scheme is given by the sequence $z_{i+1}=z_{i}(1-\epsilon)$
, where $\epsilon$ is a small parameter. Physically, for a proper
small value of $\epsilon$, this scheme implies that the ALP-photon
mixing propagation distance $l\simeq\epsilon\mathcal{H}^{-1}$ remains
well within the comoving Hubble radius $\mathcal{H}^{-1}$. Throughout
the paper we set $\epsilon=0.1$. In each interval $z\in[z_{i},z_{i+1}]$,
we approximate $\Delta(z)$ by a linear expansion centered at $z_{c}=(z_{i}+z_{i+1})/2$
as 
\begin{eqnarray}
	\Delta(z)\simeq\Delta(z_{c})+\Delta'(z_{c}) & (z-z_{c}),\label{eq:Apen-linear-exp}
\end{eqnarray}
where $\Delta(z)=\Delta_{pl}(z)-\Delta_{a}(z)$, $\Delta'(z)=\Delta'_{pl}(z)-\Delta'_{a}(z)$
and prime is derivative with respect to the redshift. Then we substitute
$dl=dz/H$ and approximate Eq. \ref{eq:Apen-P} as
\begin{eqnarray}
	\mathcal{P}(z_{i},z_{i+1}) & \simeq & \frac{3}{8(2\pi)^{2}}\frac{(1+z_{c})^{2}g_{a\gamma}^{2}}{H_{c}^{2}}\int k^{2}P_{B}dk\int_{-1}^{+1}d(\cos\theta)\int_{z_{i}}^{z_{i+1}}dz_{1}\int_{z_{i}}^{z_{i+1}}dz_{2}\nonumber \\
	&  & \times\exp\left(ik\cos\theta\frac{1}{H_{c}}\left(z_{1}-z_{2}\right)+i\int_{z_{1}}^{z_{2}}\frac{1}{\left(1+z_{c}\right)H_{c}}dz_{3}(\Delta_{pl}(z_{3})-\Delta_{a}(z_{3}))\right)\nonumber \\
	& \simeq & \frac{3}{8(2\pi)^{2}}\frac{(1+z_{c})^{2}g_{a\gamma}^{2}}{H_{c}^{2}}\int k^{2}P_{B}dk\int_{-1}^{+1}d(\cos\theta)\int_{z_{i}}^{z_{i+1}}dz_{1}\int_{z_{i}}^{z_{i+1}}dz_{2}\nonumber \\
	&  & \times\exp\left(i(z_{2}-z_{1})\left(\frac{\Delta'(z_{c})}{2\left(1+z_{c}\right)H_{c}}(z_{2}+z_{1}-2z_{c})+\frac{k\cos\theta}{H_{c}}\right)\right)\nonumber \\
	& = & \frac{3\pi}{16(2\pi)^{2}}\frac{\left(1+z_{c}\right)^{3}g_{a\gamma}^{2}}{H_{c}\left|\Delta'(z_{c})\right|}\int k^{2}P_{B}dk\int_{-1}^{+1}d(\cos\theta)\nonumber \\
	&  & \times\left\{ \textrm{Erf}\left(\left(-1\right)^{1/4}\sqrt{\frac{\left|\Delta'(z_{c})\right|}{2\left(1+z_{c}\right)H_{c}}}\left(z_{c}-\frac{\left(1+z_{c}\right)k\cos\theta}{\Delta'(z_{c})}-z_{i+1}\right)\right)\right.\nonumber \\
	&  & \left.-\textrm{Erf}\left(\left(-1\right)^{1/4}\sqrt{\frac{\left|\Delta'(z_{c})\right|}{2\left(1+z_{c}\right)H_{c}}}\left(z_{c}-\frac{\left(1+z_{c}\right)k\cos\theta}{\Delta'(z_{c})}-z_{i}\right)\right)\right\} \nonumber \\
	&  & \times\left\{ -\textrm{Erf}\left(\left(-1\right)^{3/4}\sqrt{\frac{\left|\Delta'(z_{c})\right|}{2\left(1+z_{c}\right)H_{c}}}\left(z_{c}-\frac{\left(1+z_{c}\right)k\cos\theta}{\Delta'(z_{c})}-z_{i+1}\right)\right)\right.\nonumber \\
	&  & \left.+\textrm{Erf}\left(\left(-1\right)^{3/4}\sqrt{\frac{\left|\Delta'(z_{c})\right|}{2\left(1+z_{c}\right)H_{c}}}\left(z_{c}-\frac{\left(1+z_{c}\right)k\cos\theta}{\Delta'(z_{c})}-z_{i}\right)\right)\right\} ,
\end{eqnarray}
where $H_{c}=H(z_{c})${\small{} and error function $\textrm{Erf}(x)=\frac{2}{\sqrt{\pi}}\int_{0}^{x}e^{-t^{2}}dt$.
	Thanks to the step-like shape of the error function, one can infer
	the integral over $\cos\theta$ from its geometric meaning and find
	\begin{eqnarray}
		\mathcal{P}(z_{i},z_{i+1}) & = & \frac{3\pi}{4(2\pi)^{2}}\frac{(1+z_{c})^{3}g_{a\gamma}^{2}}{H_{c}|\Delta'(z_{c})|}\int dkk^{2}P_{B}(k)W(t_{1},t_{2};k)\label{eq:Apen-P-general}
	\end{eqnarray}
	with 
	\begin{eqnarray}
		t_{1} & = & \frac{\left(z_{c}-z_{f}\right)\Delta'(z_{c})-\Delta(z_{c})}{\left(1+z_{c}\right)k},\,\,\,\,\, \hfill t_{2}=\frac{\left(z_{c}-z_{i}\right)\Delta'(z_{c})-\Delta(z_{c})}{\left(1+z_{c}\right)k},
	\end{eqnarray}
	and }
\begin{eqnarray}
	W(t_{1},t_{2};k) & = & \begin{cases}
		|t_{1}-t_{2}|, & |t_{1}|<1\quad|t_{2}|<1\\
		|t_{1}+1|, & |t_{1}|<1\quad t_{2}<-1\\
		|t_{1}-1|, & |t_{1}|<1\quad t_{2}>1\\
		|t_{2}+1|, & t_{1}<-1\quad|t_{2}|<1\\
		|t_{2}-1|, & t_{1}>1\quad|t_{2}|<1\\
		2, & t_{1}<-1\quad t_{2}>1\quad or\quad t_{1}>1\quad t_{2}<-1\\
		\simeq0, & \textrm{others}
	\end{cases}.\label{eq:Apen-W}
\end{eqnarray}
Note that Eq. \ref{eq:Apen-P-general} holds under the linear expansion
of $\Delta$-term (see Eq. \ref{eq:Apen-linear-exp}) in $[z_{i},z_{i+1}]$,
which is valid given our choice of  the discritization
parmeter $\epsilon=0.1$ and parameter space considered in this paper.

Let us apply our main formula Eq. \ref{eq:Apen-P-general} to analyze
two limiting cases. The first case considers a mass-equal resonance
in presence of the stochastic magnetic field. This resonance arises
at a redshift $z_{res}$ when the mass-equal condition $\Delta(z_{res})=m_{\gamma}(z_{res})-m_{a}(z_{res})=0$
is satisfied, producing a peak in the oscillation length profile (see
Fig. 1 in the main text). Assuming a monochromatic spectrum $P_{B}(k)=\pi^{2}B_{0}^{2}\delta\left(k-k_{B}\right)/k_{B}^{2}$
and a constant re-ionization fraction $X_{e}=10^{-4}$, from Eq. \ref{eq:Apen-P-general}
we derive the conversion probability over $[z_{i}=z_{res}(1+\epsilon/2),z_{i+1}=z_{res}(1-\epsilon/2)]$
as
\begin{eqnarray}
	\mathcal{P} & \simeq & 8.6*10^{-4}\left(\frac{g_{a\gamma}}{10^{-12}\textrm{GeV}^{-1}}\right)^{2}\left(\frac{B_{0}}{\textrm{nGs}}\right)^{2}\left(\frac{\omega_{0}}{\textrm{GHz}}\right)\left(\frac{m_{a}}{10^{-12}\textrm{eV}}\right)^{1/3}\textrm{Min}\left(\epsilon z_{res}\left(\frac{\textrm{GHz}}{\omega_{0}}\right)\left(\frac{\lambda_{B}}{\textrm{kpc}}\right),1\right).
\end{eqnarray}
One can see that the front factor is of similar magnitude and scaling
on each variable as those obtained in Landau-Zener approximation \citep{Marsh:2021ajy,Mirizzi:2009nq,Choi:2019jwx,Mondino:2024rif,Tashiro:2013yea}
or the same spirit but so called saddle-point approach \citep{Chen:2013gva}.
Interestingly, our results introduce a Minimum function that highlight
a novel suppression effect, which arises from the stochastic nature
of the magnetic background. In fact, 
the $\omega_{0}^{-2}$ scaling of the brightness temperature shown in Fig. 2 in the main text transits to $\omega_{0}^{-3}$ at the higher
frequency $\omega_{0}\gtrsim100$GHz.
Such a suppression is absent in commonly
used domain-like model in literature. In domain-like model, the magnetic
field is treated as constant over a domain patch in size of correlation
length $\lambda_{B}$. Thus the Landau-Zener approximation is only
valid when domain size exceeds the resonance width $W_{res}$, namely
$\lambda_{B}\gtrsim W_{res}$. Our approach, however, is out of this
limit and can extend to the case with small $\lambda_{B}$, where
a suppression is revealed as a consequence of stochastic magnetic
field and narrowness of the mass-equal resonance.

The second case corresponds to a rather slowly varying oscillation
length scale (in physical frame) with $\Delta'<<1$. Again, we assume
a monochromatic spectrum for stochastic magnetic field. When the condition
$-1<t_{1}\simeq t_{2}=\frac{-\Delta(z_{c})}{(1+z_{c})k_{B}}<1$
is met, we find $W(t_{1},t_{2})=\epsilon|\Delta'(z_{c})|/k_{B}$ and
hence the conversion probability reads
\begin{eqnarray}
	\mathcal{P} & \simeq & \frac{3}{32}z_{c}^{2}g_{a\gamma}^{2}B_{0}^{2}\lambda_{B}\triangle l,
\end{eqnarray}
where $\triangle l\simeq\epsilon z_{c}/H_{c}$ is approximately the comoving
distance during $[z_{i},z_{i+1}]$. Under the condition $\left|\frac{\Delta(z_{c})}{(1+z_{c})k_{B}}\right|<1$,
the conversion probability shows a linear growth with distance. This
condition can also be expressed as $\pi\lambda_{B}<l_{osc}$ given
the definition of comoving oscillation length $l_{osc}(z)=2\frac{1+z}{\Delta(z)}$.
Such a resonant condition associated with the stochastic magnetic
field has also been identified in Ref. \citep{Addazi:2024kbq}.

In this paper, we study two representative cases with $m_{a}=3\times10^{-14}\textrm{eV}$
and $m_{a}=3\times10^{-13}\textrm{eV}$. As shown in Fig. 1 in the main text,
the oscillation curve consists of the peak at mass-equal resonance
and relatively slow-varying region.We employ the semi-steady approximation
across the entire post recombination epoch with redshift sequence
$[z_{1},z_{2},...,z_{N}]$ and have verified that the linear expansion
condition holds as mentioned above. The total conversion probability
from recombination up to $z_{i}$ is simply the sum $\mathcal{P}^{tot}(z_{i})=\sum_{j=1}^{j=i}\mathcal{P}(z_{j})$.
By interpolating $\mathcal{P}^{tot}(z_{i})$ $(i=1,2,...)$ over the
redshift sequence, one can obtain the total probability $\mathcal{P}^{tot}(z)$
from recombination to any value of $z$.

\section*{Appendix B}

The transition from ALPs to photons results in an excess radiation
proportional to the conversion probability, specifically \citep{Ejlli:2019bqj,Fujita:2020rdx,Domcke:2020yzq}
\begin{eqnarray}
	\Omega_{\gamma}(\omega,T(z)) & = & \Omega_{ALP}(\omega,T(z))\mathcal{P}^{tot}(\omega,z).\label{eq:Apen-energy-conver}
\end{eqnarray}
Here $\Omega_{\gamma}(\omega,T(z))$ and $\Omega_{ALP}(\omega,T(z))$
respectively represent the energy density per logarithmic frequency
interval of photons and ALPs at temperature $T$ (equivalently at
redshift $z=T/T_{0}-1$). Working within the Rayleigh-Jeans frequency
regime, we translate the radiation energy density to the brightness
temperature 
\begin{eqnarray}
	T_{b}(\omega,T) & = & \frac{2\pi^{2}}{\omega^{2}}I(\omega,T)=\frac{2\pi^{2}}{\omega^{2}}\frac{\omega^{3}}{2\pi^{2}}f_{\gamma}(\omega,T)=\frac{2\pi^{2}}{\omega^{2}}\frac{\omega^{3}}{2\pi^{2}}\frac{\Omega_{\gamma}(\omega,T)\pi^{2}\rho_{c}(T)}{\omega^{4}},\label{eq:Apen-Tb-photon}
\end{eqnarray}
where the spectral intensity is $I(\omega,T)=\frac{\omega^{3}}{2\pi^{2}}f_{\gamma}(\omega,T)$
with distribution function $f_{\gamma}=\left(\exp\left(\omega/T\right)-1\right)^{-1}$
and $\Omega_{\gamma}(\omega,T)=\frac{\omega^{4}f_{\gamma}(\omega,T)}{\pi^{2}\rho_{c}}$.
The universe critical density $\rho_{c}(T)$ can be expressed in terms
of the energy density $\boldsymbol{\Omega}_{\gamma}(T)=\int d\lg\omega\Omega_{\gamma}(\omega,T)=\frac{\pi^{2}T^{4}}{15\rho_{c}}$,
thus Eq. \ref{eq:Apen-Tb-photon} becomes
\begin{eqnarray}
	T_{b}(\omega,T) & = & \frac{\pi^{4}}{15}\left(\frac{T}{\omega}\right)^{3}\frac{\Omega_{\gamma}(\omega,T)}{\boldsymbol{\Omega}_{\gamma}(T)}T=\frac{\pi^{4}}{15}\frac{T^{4}}{\omega^{3}}\frac{\Omega_{ALP}(\omega,T)}{\boldsymbol{\Omega}_{\gamma}(T)}\mathcal{P}^{tot}(\omega,z),\label{eq:Apen-Tb-P}
\end{eqnarray}
where Eq. \ref{eq:Apen-energy-conver} is used in the second equality.

Considering the scaling relations $T,\omega\propto a^{-1}$ and $\Omega_{ALP}(\omega,T),\Omega_{\gamma}(T)\propto a^{-4}$,
we obtain the brightness temperature converted from ALPs at present
day as
\begin{eqnarray}
	T_{b}(\omega_{0}) & = & \frac{\pi^{4}}{15}\frac{T_{0}^{4}}{\omega_{0}^{3}}\frac{\Omega_{ALP0}(\omega_{0})}{\boldsymbol{\Omega}_{\gamma0}}\mathcal{P}^{tot}(\omega_{0}).\label{eq:Apen-Tb-ARCADE}
\end{eqnarray}
The subscript ``$0$'' denotes the present value and $\mathcal{P}^{tot}(\omega_{0})$
is the total conversion probability accumulated from recombination
until now. We use Eq. \ref{eq:Apen-Tb-ARCADE} to calculate the excess
radiation from the ALPs and compare it with observational excess by
ARCADE-2 and other lower-frequency measurements. To assess the impact
on the 21cm global signal, one needs to count the excessive photon
at a 1.4GHz rest frequency converted from the ALPs at any given redshift.
Specifically, the additional brightness temperature from ALP conversion
is given by
\begin{eqnarray}
	T_{b}(z) & = & \frac{\pi^{4}}{15}\left(\frac{T_{0}}{\widetilde{\omega}}\right)^{3}\frac{\Omega_{ALP}(\widetilde{\omega})}{\boldsymbol{\Omega}_{\gamma}}\mathcal{P}^{tot}(\widetilde{\omega},z)T_{0}(1+z),\label{eq:Apen-Tb-EDGES}
\end{eqnarray}
where $\widetilde{\omega}=2\pi\times1.4\textrm{GHz}/(1+z)$ and the
current CMB temperate is $T_{0}=3.57\times10^{11}\textrm{Hz}$. Given
a certain redshift $z$, the $\mathcal{P}^{tot}(\widetilde{\omega},z)$
represents the accumulated probability of ALP-to-photon conversion
at frequency $\widetilde{\omega}$ from recombination up to that redshift.
Including the CMB radiation, the total brightness temperature relevant
to 21cm physics at given redshift is $T_{b}(z)+T_{0}(1+z)$, which
is used in simulation and compared to EDGES.

Several points need to be stressed. First,  we assume  a frequency-independent
$\Omega_{ALP}(\omega)$ for the abundance of ALPs in this work. Consequently,
the ALP energy density is $\boldsymbol{\Omega}_{ALP}=\int d\lg\omega\Omega_{ALP}(\omega)\simeq\Omega_{ALP}(\omega)$
due to the logarithmic integral manner. Thus the ratio of ALP to photon
energy density, $\gamma=\Omega_{ALP}(\omega)/\boldsymbol{\Omega}_{\gamma}\simeq\boldsymbol{\Omega}_{ALP}/\boldsymbol{\Omega}_{\gamma}=\boldsymbol{\Omega}_{ALP0}/\boldsymbol{\Omega}_{\gamma0}$,
is a frequency-independent parameter fixed at the present time. This
ratio is constrained by $\gamma=0.23\triangle N_{\textrm{eff}}\lesssim0.06$
with the extra effective number of neutrino species $\triangle N_{\textrm{eff}}\lesssim0.23$
recently reported in \textit{Planck} \citep{Planck:2015zrl}.
Second point is regarding the origin of the ALP production.  In our relevant parameter space, high-frequency ALPs are relativistic, effectively contributing as dark radiation. For their abundance to saturate the upper limit set by the $\triangle N_{\textrm{eff}}$ constraint, these ALPs must decouple from the primordial plasma at an energy scale of approximately $0.1\,\textrm{GeV}$ \citep{Caloni:2022uya}. Furthermore, within this same parameter range, ALPs at lower frequencies may exist as non-relativistic particles. With a significantly suppressed misalignment angle, $\theta \sim 0.01$, these non-relativistic ALPs could oscillate near the potential minimum, behaving as cold dark matter \citep{Arias:2012az}. The last point to mention is that the parameter combination $g_{a\gamma}B\lesssim5\times10^{-14}{\rm GeV^{-1}nGs}$
in our model's viable region complies with CMB distortion constraints
from the photon to ALP resonant conversion \citep{Mirizzi:2009nq}.

We further comment on a well-motivated theoretical framework for ALPs with mass $m_a \sim 10^{-14}$--$10^{-12}\,\mathrm{eV}$ and ALP-photon coupling $g_{a\gamma} \sim 10^{-23}$--$10^{-21}\,\mathrm{GeV}^{-1}$ beyond the Standard Model. 
In particular, ALPs arising from flavor symmetry considerations and neutrino mass generation mechanisms naturally populate the relevant mass-coupling region of our model. For example, 
it was proposed a $SU(3)_{H}\times  U(1)_H$ model as a minimal framework for realistic quantum flavourdynamics, where $SU(3)_{H}$ is a spontaneously broken local family symmetry, and $U(1)_H$ is its associated global symmetry \cite{Berezhiani:1990jj,Berezhiani:1989fp}. This model effectively reproduces the mass hierarchy and mixing patterns of quarks and leptons. Key predictions include the existence of neutrino Majorana masses with a regular hierarchy, the presence of a familon that simultaneously acts as an invisible axion (or arion) and a Majoron, and a specific correlation between neutrino lifetimes and familon decays. Consequently, the model provides a unified foundation for all primary forms of dark matter relevant to the theory of large-scale structure in the Universe.
In this model and similars, the PNGBs, which are ALPs, have an effective coupling with extra sterile neutrinos 
scaling as
$g_{a NN}\sim 1/f_{a}$ where $f_{a}$ is a characteristic spontaneous symmetry breaking 
scale.
Contrary to QCD axion, PNGBs do not acquire a contribution from the QCD instantons,
i.e. ALPs can have a much lighter mass than QCD axions. 
Thus, the mass does not  (approximately) scale as $m_{a}\sim m_{\pi}^2/f_{a}$.
Moreover, a mass term for these particles can be generated
by loop corrections with sterile neutrinos (typically introduced in these models),
with a leading order scaling (neglecting subleading log-corrections)
of 
$m_{a}\sim m_{N}^{2}/f_{a}$.
Specifically, the PGB with mass $m_{a}\sim10^{-13}eV$
can be obtained with a natural choice of parameters $f_{a}\simeq 10^{12}\, {\rm GeV}$ 
and $m_{N}\simeq 10\, {\rm KeV}$ (corresponding to a possible candidate of Warm Dark Matter) 
assuming a ALP-photon coupling of $g_{a\gamma}\sim0.01f_{a}^{-1}\sim10^{-14}GeV$
compatible with the coupling range adopted in our model.
This represents a concrete realization of a model that simultaneously addresses both the flavor hierarchy problem and neutrino mass generation, while naturally accounting for the observed radio anomalies.

\section*{Appendix C}

\begin{figure}[tbp]			
	\includegraphics[scale=0.45]{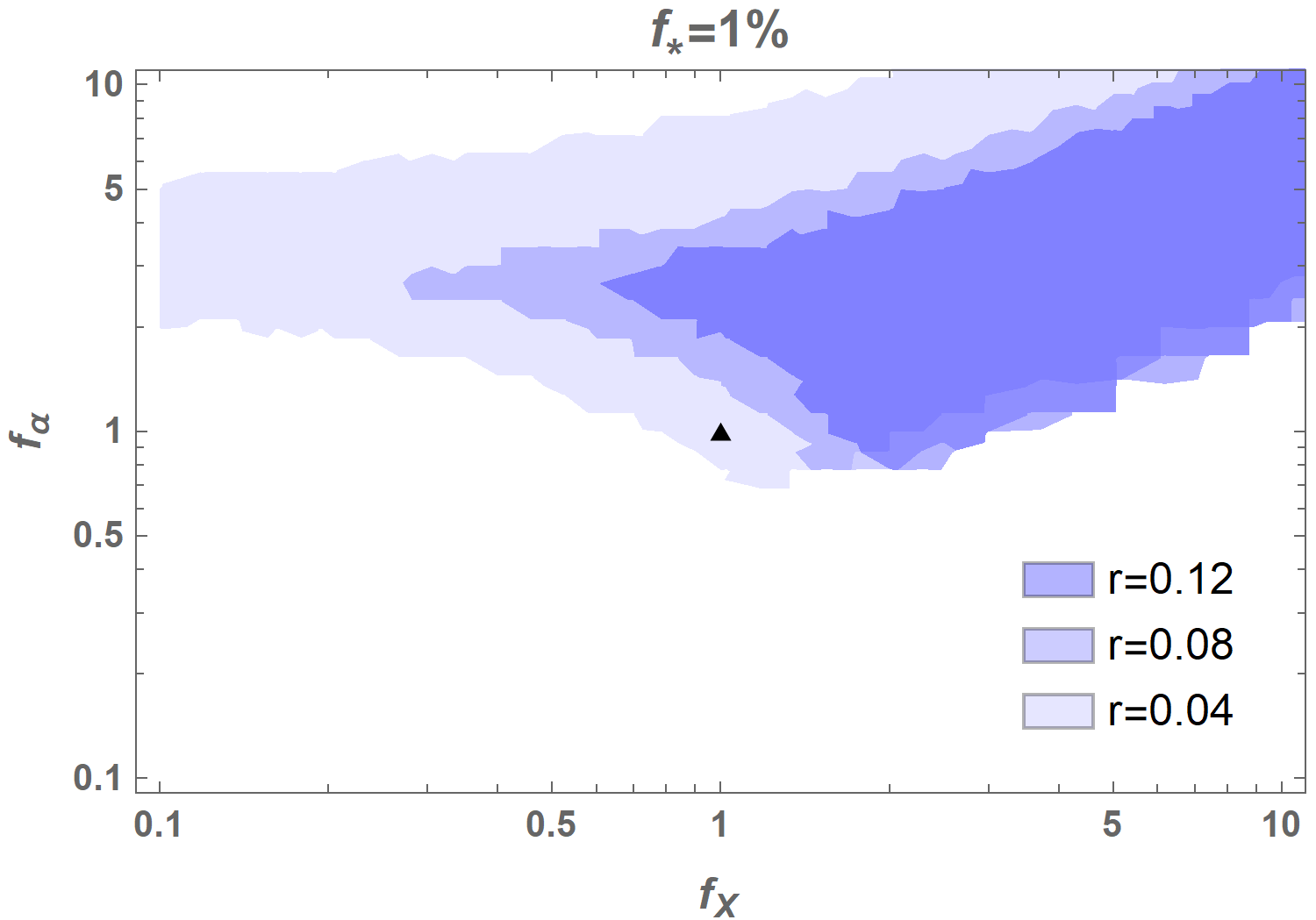} \includegraphics[scale=0.45]{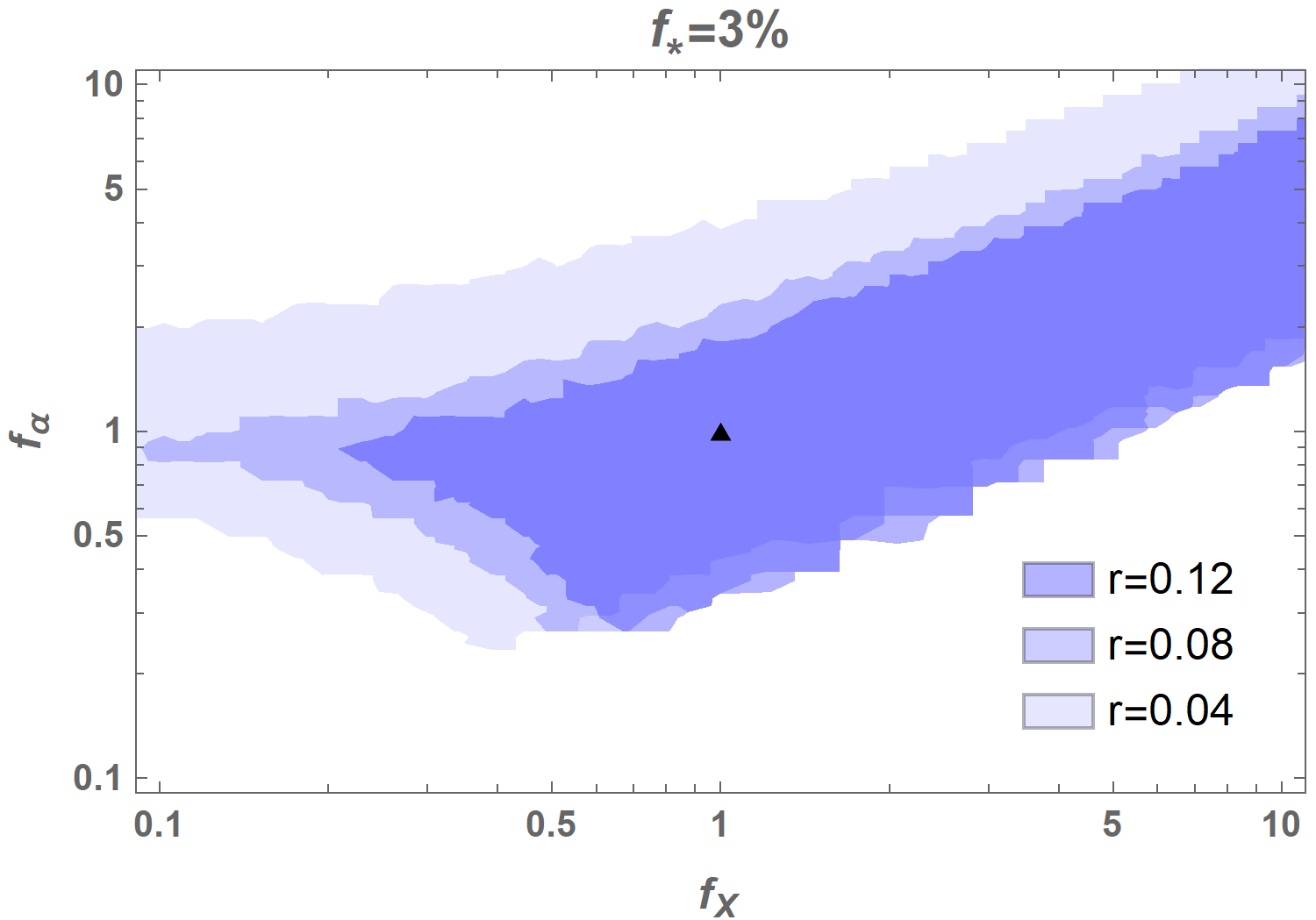}	\includegraphics[scale=0.45]{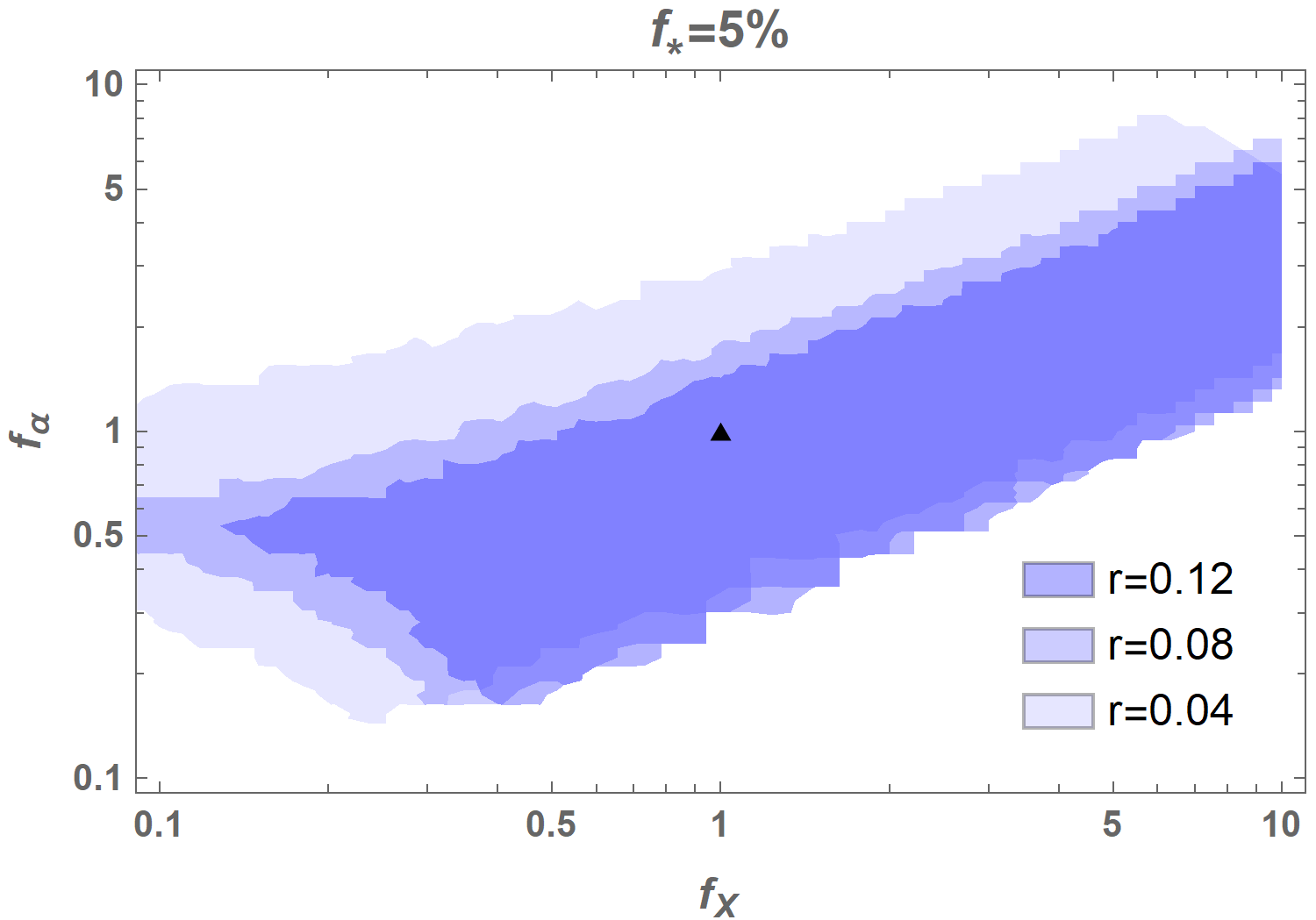}
	
	\caption{viable region in $f_{\alpha}$-$f_{X}$ space 
		that can account for the anomalous EDGES absorption signal, with   $r$ fixed to the benchmark  values adopted  in Fig. 3 in the main text. The star formation efficiency is set to $1\%$, $3\%$ and  $5\%$ from left to right. The black dot denotes the  fiducial case $f_{\alpha}=f_{X}=1$.}
	
	\label{fig-falphafX}
\end{figure}

In 21cm physics, the heating process driven by Lyman-$\alpha$ and X-ray flux are determined by  normalized emissivity factors $f_{\alpha}$ and  $f_{X}$ \citep{Pritchard:2011xb}. In Fig. 3 in the main text we adopt the fiducial value $f_{\alpha}=f_{X}=1$  to illustrate the impact of axion-photon mixing  on the 21cm global signal. 
In fact, the values of  $f_{\alpha}$ and  $f_{X}$ are highly model-dependent, varying significantly with different assumptions regarding the star formation and its dynamics. 
The impact of varying $f_{\alpha}$ and $f_{X}$ on  21cm global signal and power spectrum  have been explored in Refs. \citep{Pritchard:2008da,Fialkov:2019vnb}. 
Motivated  by this, we also investigate a broader range of  $f_{\alpha}$ and $f_{X}$ values beyond  the fiducial one to examine their  affects  within our model. 
We simulate the 21cm global signal as in Ref \citep{Caputo:2020avy} and 	  
set the halo virial temperature
cut at $2\times10^{4}$K. We consider range of $f_{\alpha}$ and  $f_{X}$ both from $0.1$ to $10$, alongside  three cases of star formation efficiency
at $1\%$,$3\%$ and $5\%$. 
we search for the viable parameter space  that could
account for the absorption feature observed by EDGES. 
Specifically, this viable parameter space is identified using criteria that ensure the absorption trough minimum lies within the range $15<z<20$ and $-1\textrm{K}<\triangle T_{b}^{\textrm{21cm}}<-0.3\textrm{K}$, aligning with the EDGES data at  $99\%$ C.L.	
Our results reveal a broad viable region in all  different choices of parameter $f_*$ and $r$.
Moreover, the overall shift of the viable space with respect to different $f_*$ reflect a  degeneracy between $f_{\alpha}$,  $f_{X}$ and  $f_*$, namely only the product combination $f_{\alpha}f_*$ and  $f_{X}f_*$ determine the 21cm signal.

\section*{Appendix D}

\begin{figure}[tbp]			
	\includegraphics[scale=0.45]{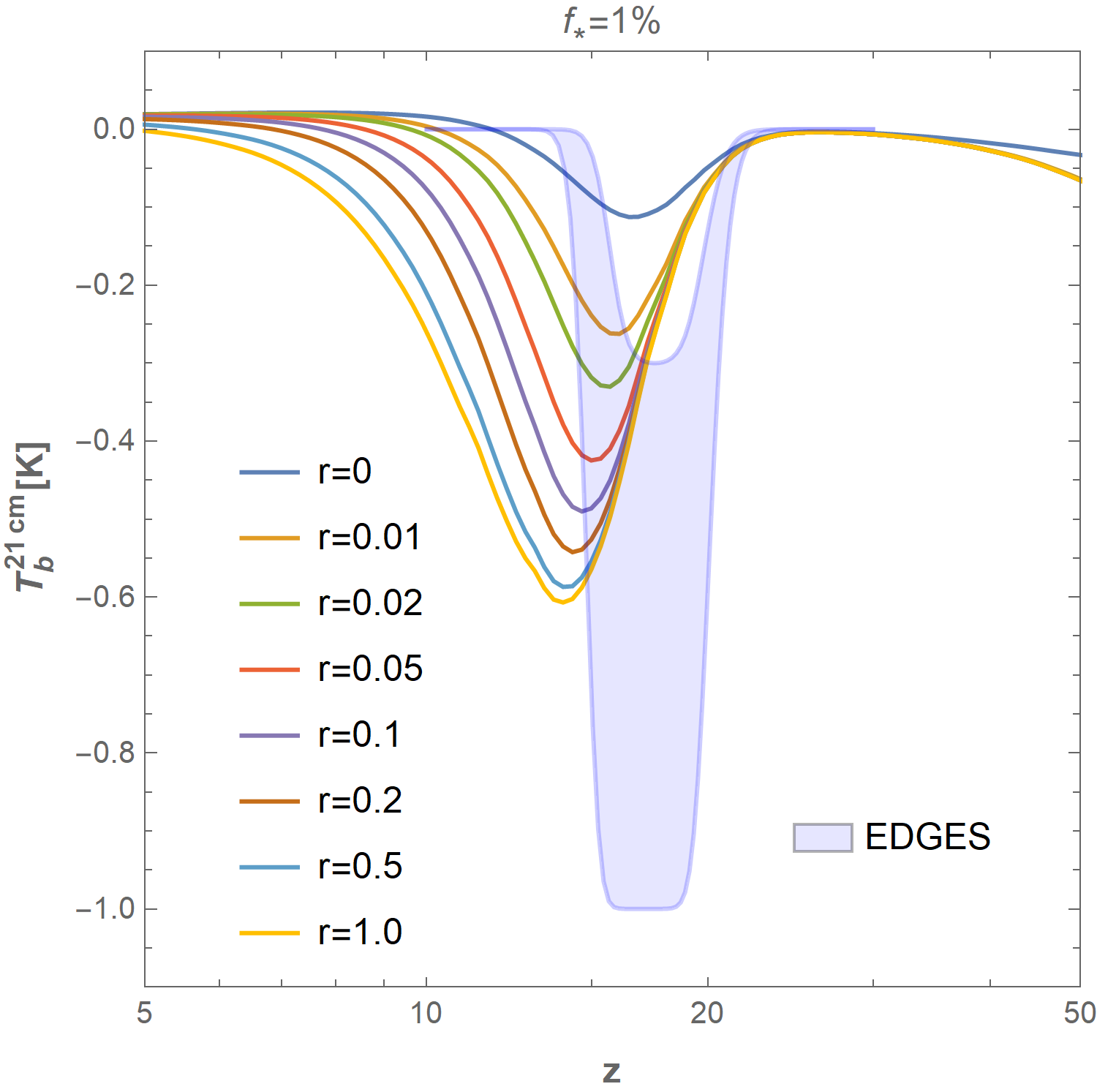}\includegraphics[scale=0.45]{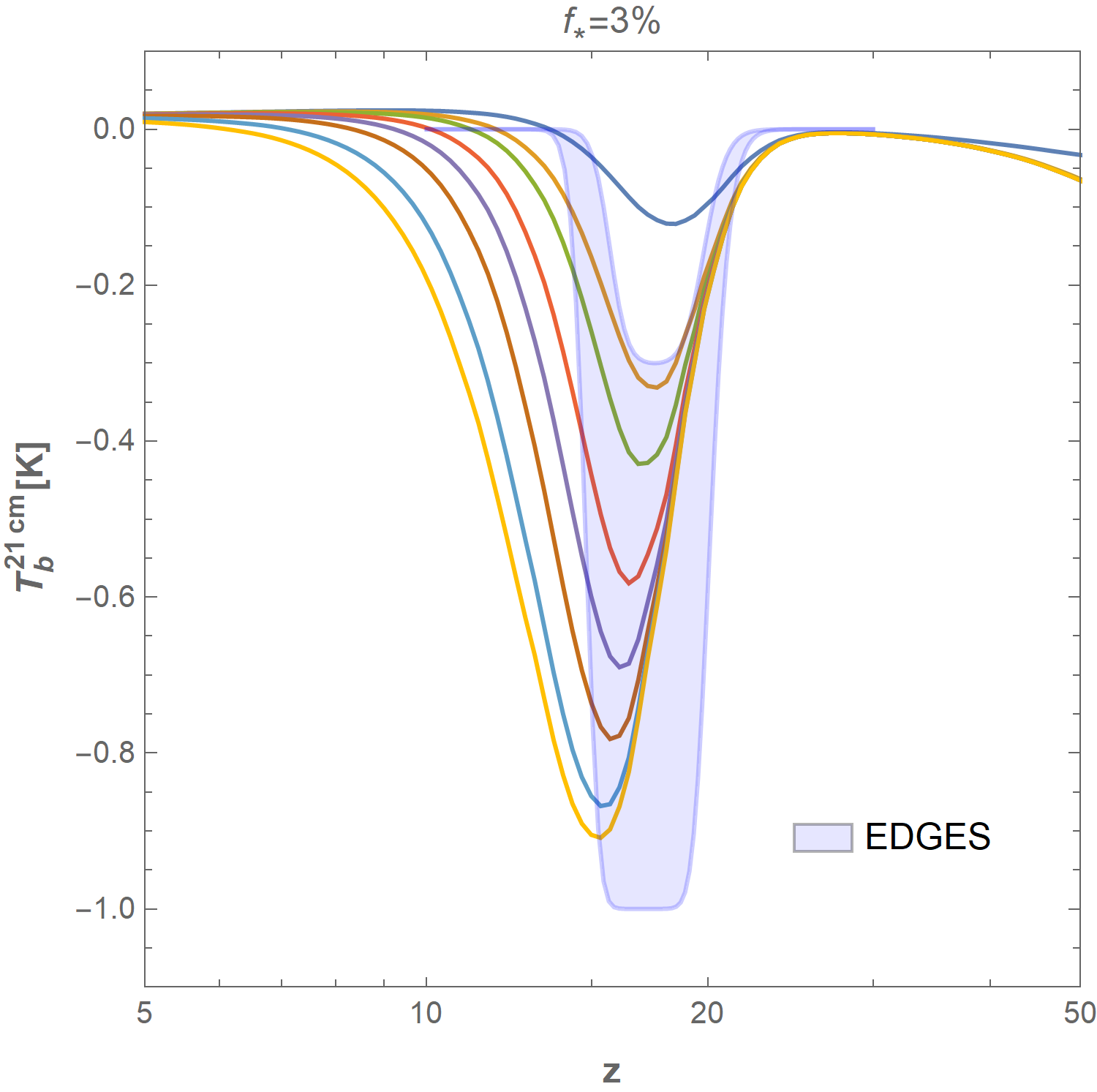}\includegraphics[scale=0.45]{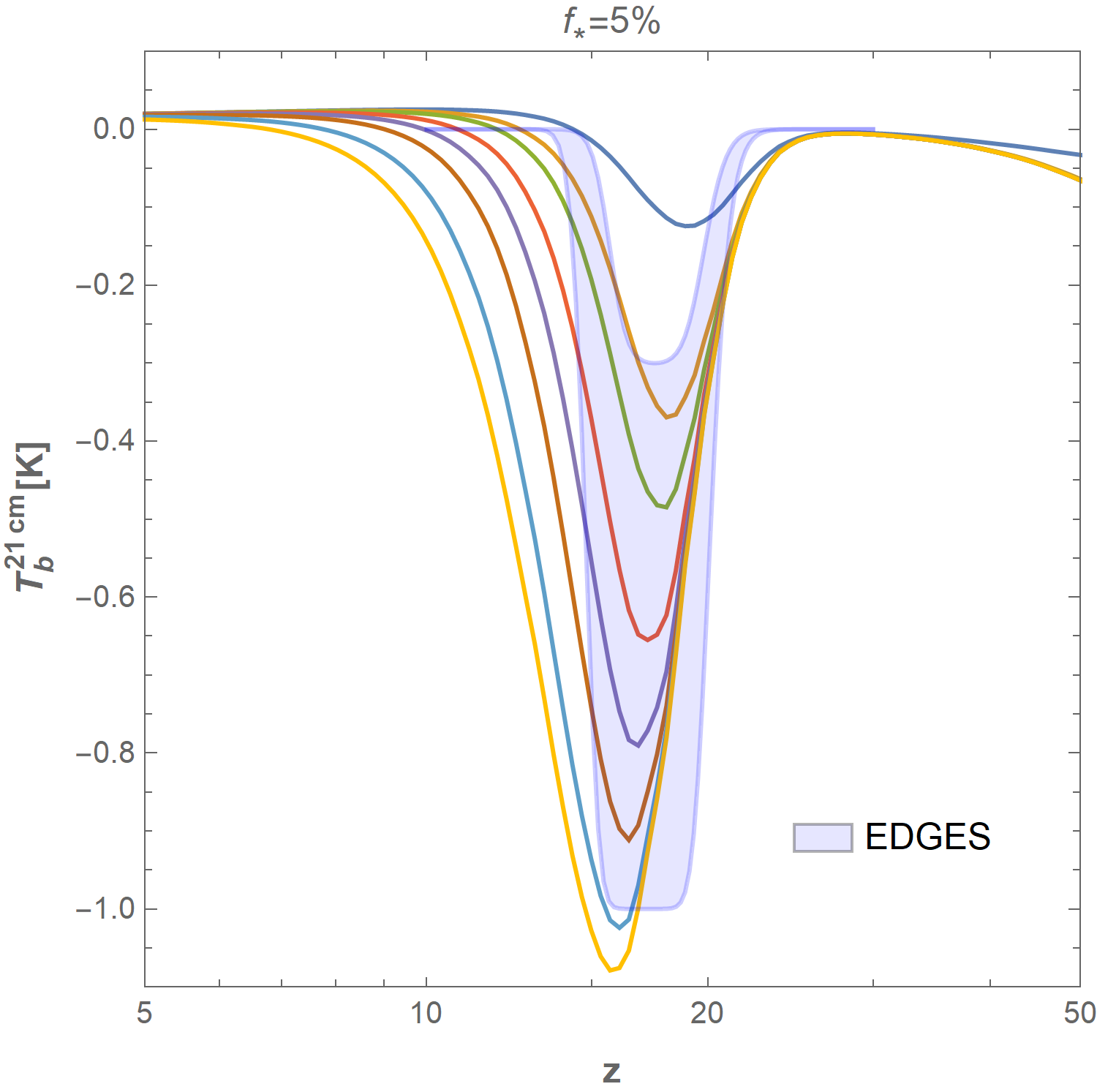}	
	\caption{The 21cm global signal under the modified radiation background for various values of the parameter 
		$r$, compared with EDGES observation. The star formation efficiency are fixed at $f_*=1\%$, $f_*=3\%$ and $f_*=5\%$ from left to right panels. }
	
	\label{fig-21cm-r}
\end{figure}

\begin{figure}[tbp]			
	\includegraphics[scale=0.7]{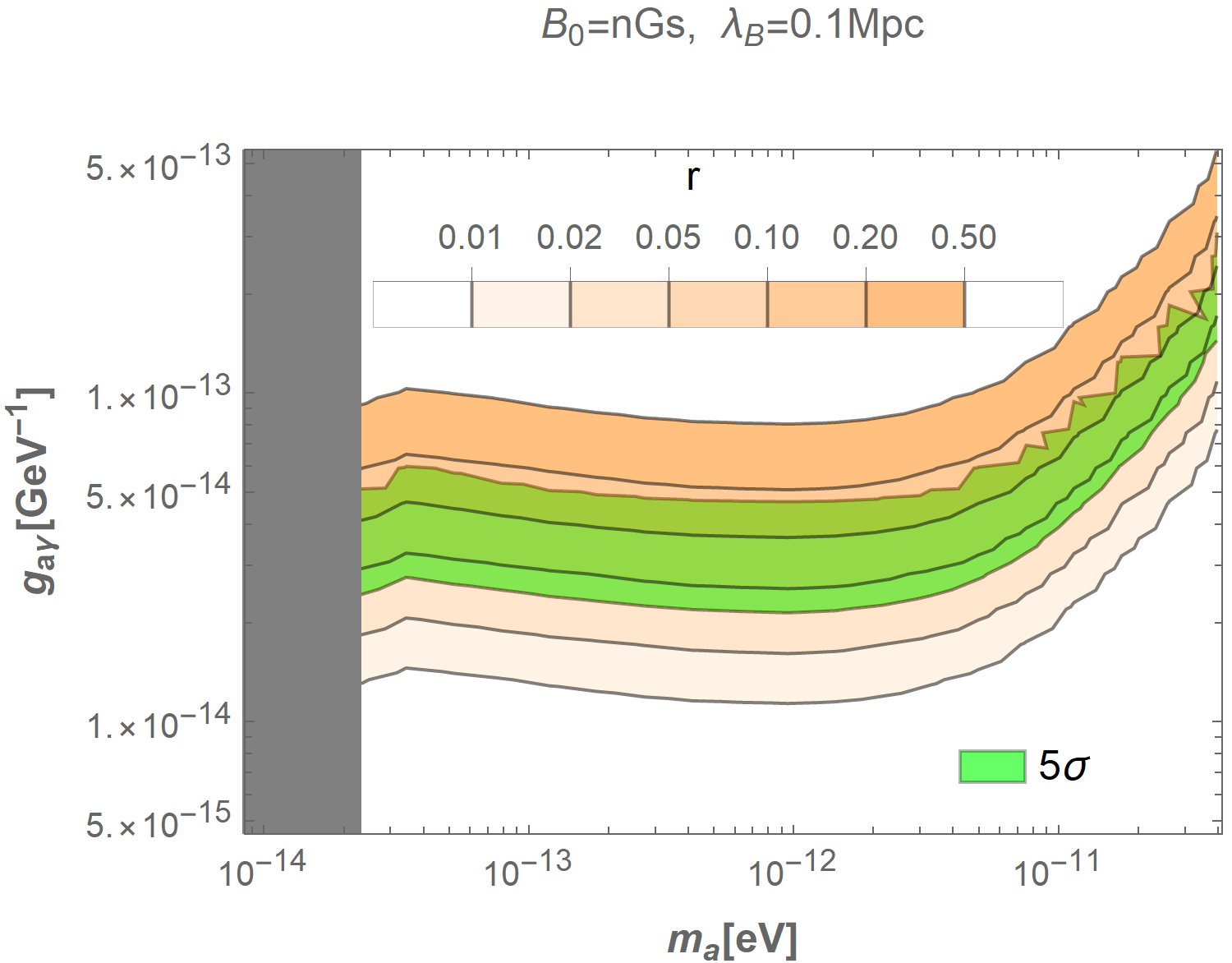}\includegraphics[scale=0.7]{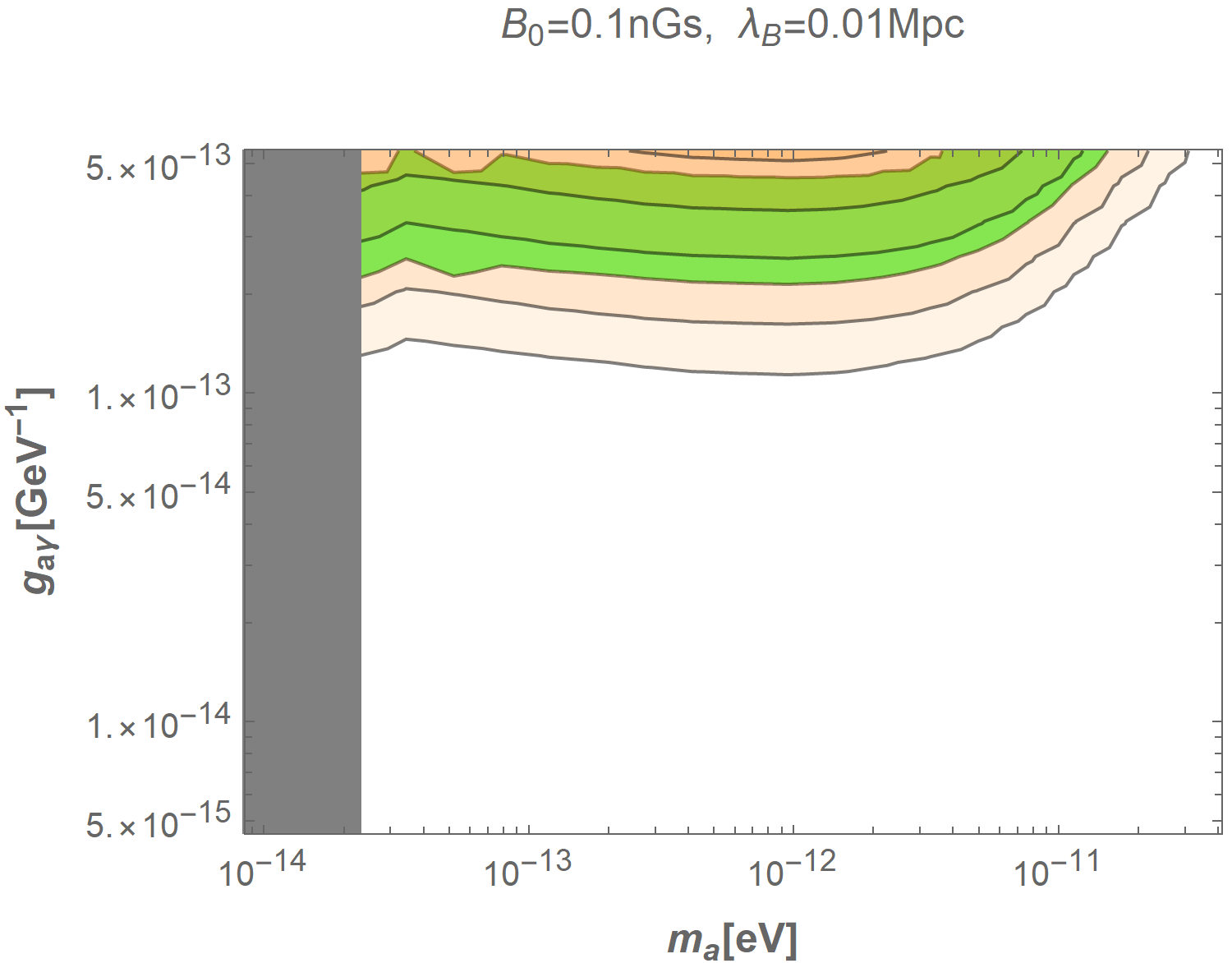}	\\
	\includegraphics[scale=0.7]{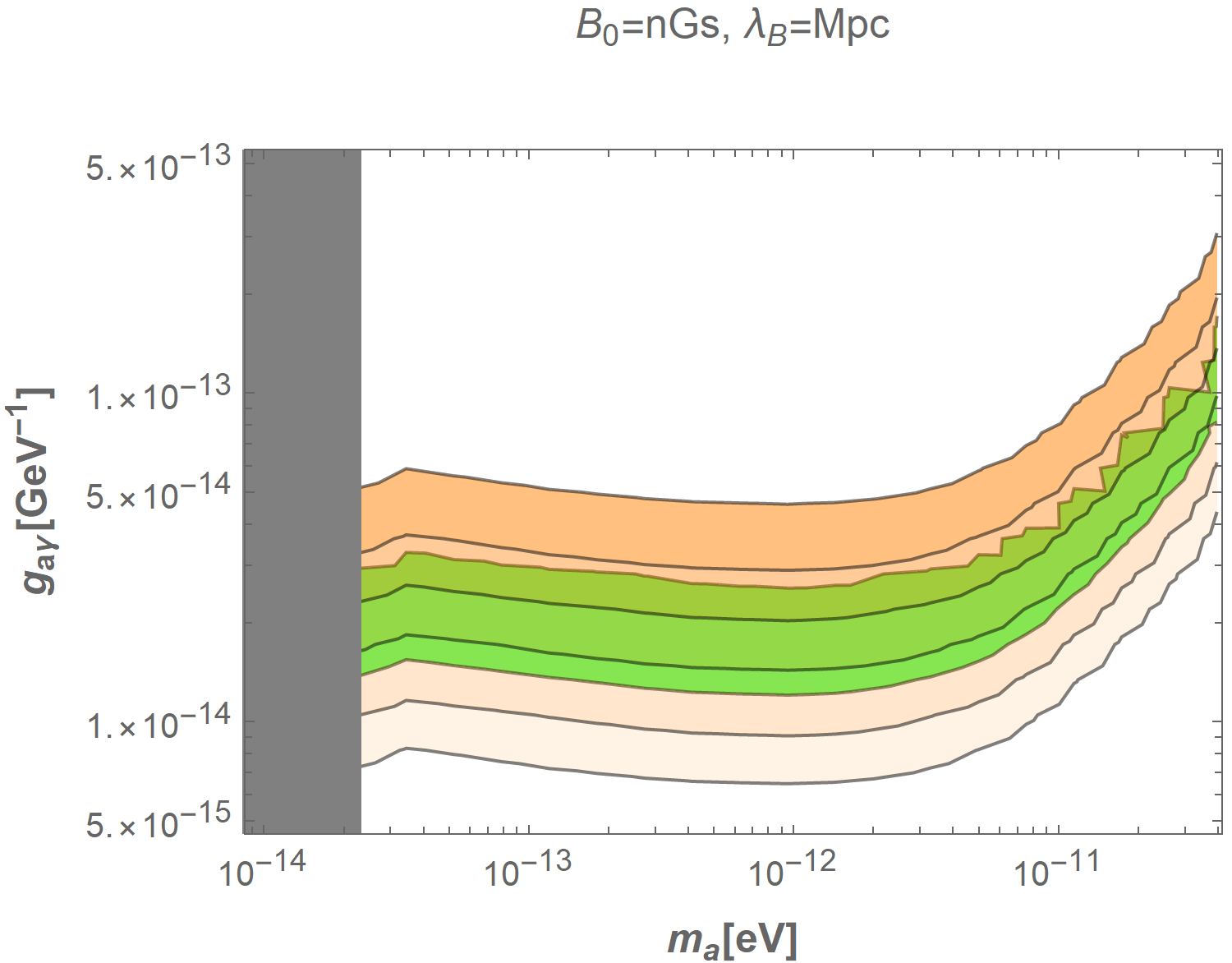}
	\includegraphics[scale=0.7]{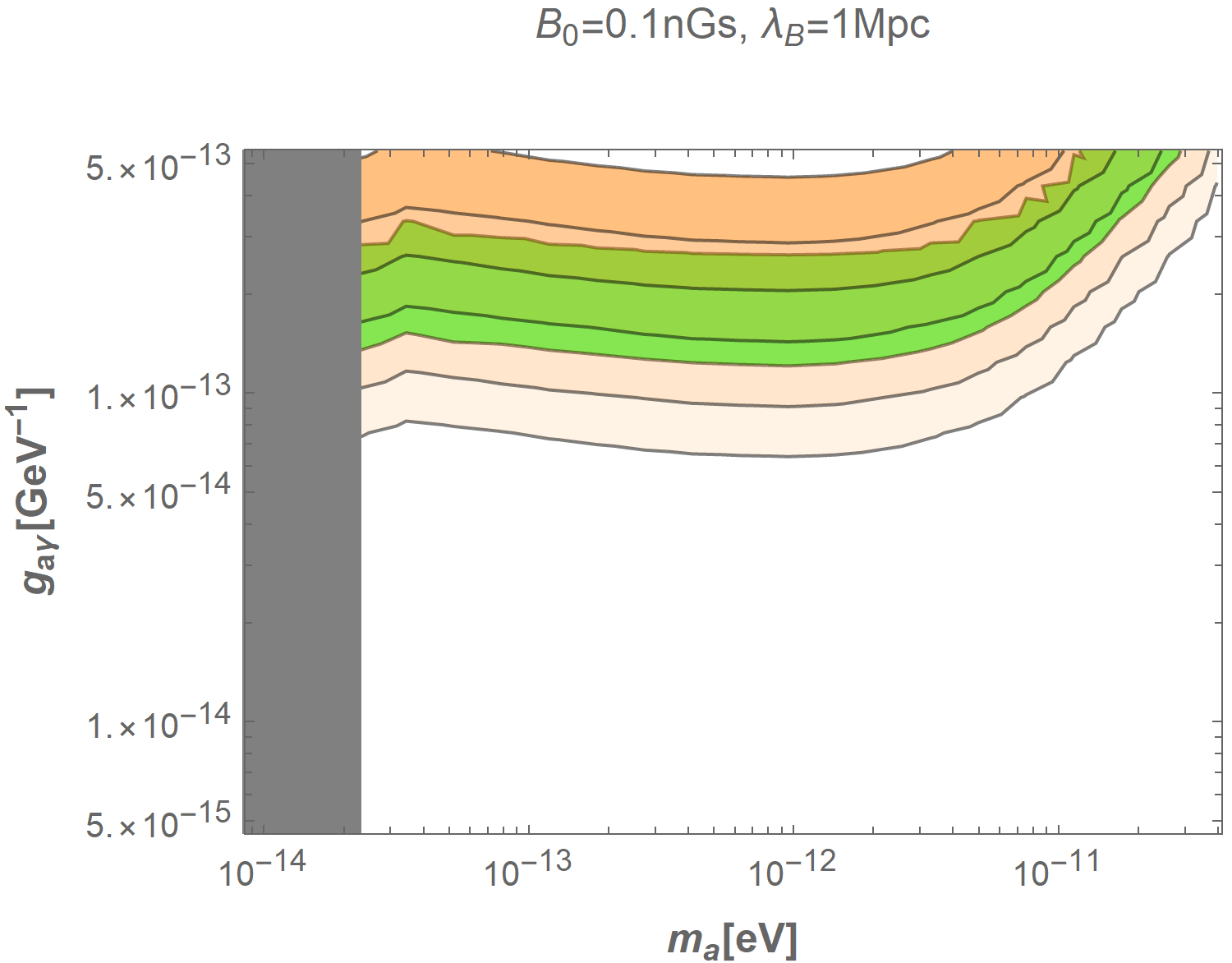}	
	
	\caption{The viable region in $g_{a\gamma}$-$m_{a}$ plane for explaining radio anomalies. The upper and lower panels correspond to the magnetic field with peaked spectrum and scale invariant  spectrum, respectively. The green region denotes the  $5\sigma$ improvement of our ALP model over the model incorporating only astrophysical synchrotron sources  (see Fig. 2 in main text). The  gray region corresponds to lower ALP mass, where the mass-equal resonance occurs at $z_{res} \lesssim 10 $, preventing the generation of a sufficiently deep 21cm absorption trough observed by EDGES. The ALP-to-photon energy density ratio is fixed at $\gamma=0.06$.  }
	
	\label{fig-g-ma}
\end{figure}		

As illustrated  in Fig.1 of the main text, the radiation background affecting 21cm in our ALP model can be parameterized as 
\begin{equation}
	T_R=T_0 (1+z)\left( 1+ r A \left( \frac{f_{obs}}{78 \mathrm{MHz}} \right)^\beta \Theta(z_{res}-z) \right)
\end{equation}
where $f_{obs}=1.4\mathrm{GHz}/(1+z)$, $A=319$, $\beta=2$ and  $T_0$ the CMB temperature today. The Heaviside function $\Theta$ function accounts for the abrupt rise in radiation at  $z_{res}$ triggered by mass-equal resonance. The parameter $r$ quantifies the intensity of the excess radiation arising from ALP-photon conversion.    
To produce a sufficiently deep 21cm absorption feature at $z\sim 17$ as detected by EDGES, the model requires $z_{res}\gtrsim 20 $ and an appropriate choice of $r$. In Fig. \ref{fig-21cm-r} we present the 21cm global signal for various values of $r$. In three cases with star formation efficiency $f_*=1,3,5\%$ considered here,  the 21cm absorption trough lies within EDGES observation for a range  $0.02 \lesssim r \lesssim 0.2$. Conservatively, we adopt this range as the minimum requirement to account for the 21cm anomaly and subsequently explore the viable parameter space in the $g_{a \gamma}$-$m$ plane that simultaneously explains the ARCADE-2 excess.

As shown in Fig. \ref{fig-g-ma}, there exists a significant overlap between $5 \sigma$ region  and the range  $0.02 \lesssim r \lesssim 0.2$, indicating a simultaneous explanation on both radio anomalies. In particular, the overlap region presents at ALP mass $m_a \gtrsim 10^{-14}\mathrm{eV}$ as required by aforementioned condition  $z_{res}\gtrsim 20$. Moreover, given the ALP-to-photon density ratio  $\gamma=0.06$, a simultaneous explanation requires an ALP-photon coupling of $g_{a \gamma} \simeq 5 \times 10^{-14}\mathrm{GeV}^{-1}$ for $B_0=\mathrm{nGs}$ and $g_{a \gamma} \simeq 5 \times 10^{-13}\mathrm{GeV}^{-1}$ for  $B_0=0.1\mathrm{nGs}$, regardless of whether the primordial magnetic field follows a peaked or scale-invariant spectrum. In fact, the radiation excess driven by ALP-photon conversion can be  schematically captured as $T_R \propto \gamma g_{a\gamma}^2 B_0^2 $. Thus a possible relaxed condition on $\gamma$ in the post-recombination era can further lower the coupling $g_{a \gamma}$ and the magnetic strength $B_0$.

\section*{Appendix E} 

\begin{figure}[tbp]			
	\includegraphics[scale=0.6]{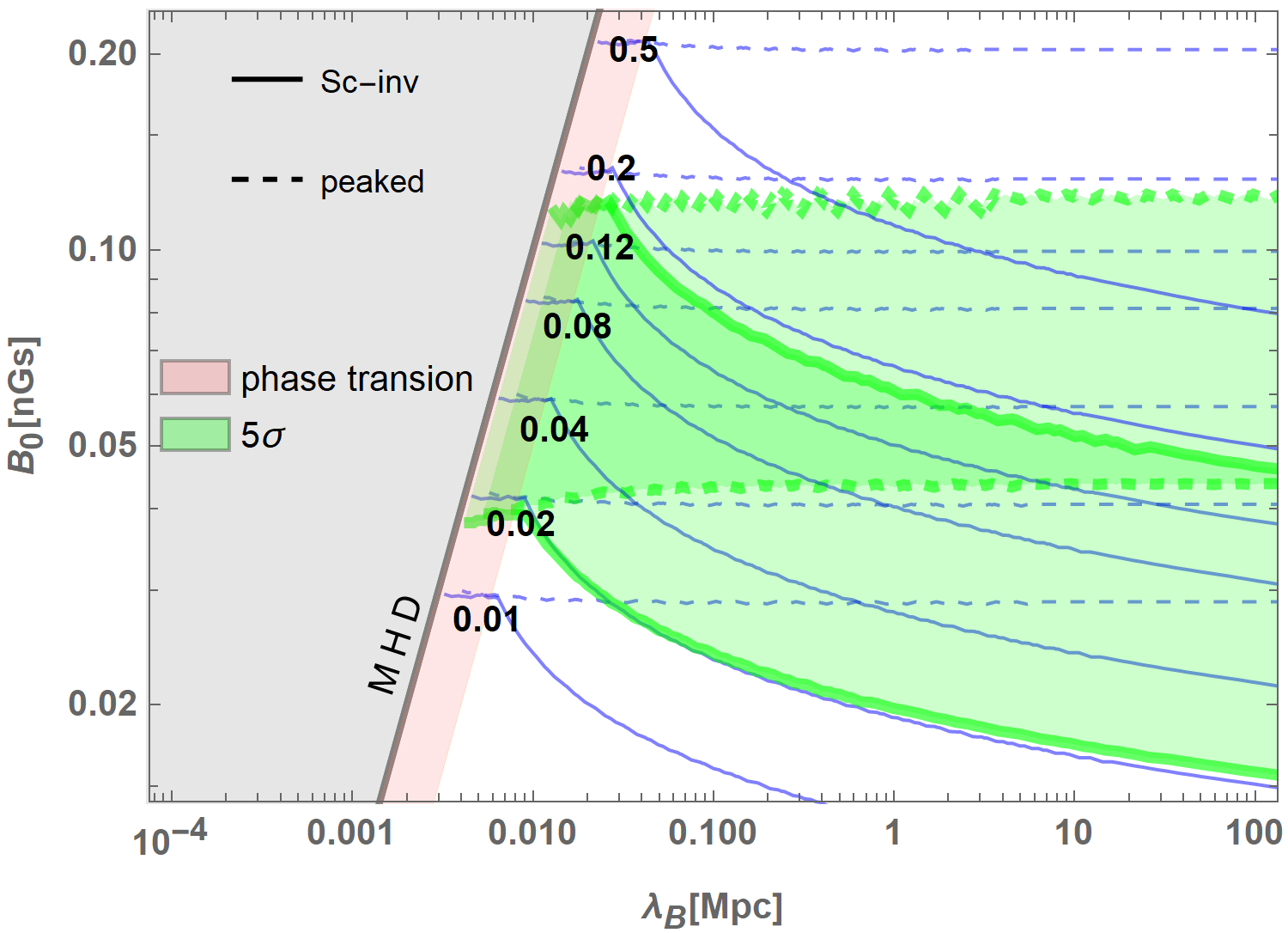}	
	
	\caption{We perform a comparative analysis between scale-invariant and peaked spectra within the white region, corresponding to the full parameter space to the right of the pink exclusion zone. The green shaded area, bounded by solid (scale-invariant) and dashed (peaked spectrum) green contours, indicates the parameter space that yields a $5\sigma$ improvement in fitting the radio excess data within our ALP framework. 
		The corresponding $r$-parameter curves are shown as solid (scale-invariant) and dashed (peaked) blue lines. The range $0.02 \lesssim r \lesssim 0.1$ consistently explains the 21~cm absorption anomaly for both spectral cases (see Fig.~2 in the main text for details). The ALP mass is fixed at $m_{a}=3\times10^{-14}$eV and ALP-to-photon energy density ratio at $\gamma=0.06$.}
	
	\label{fig-peaked-All}
\end{figure}

In this appendix, we briefly comment on the magnetic spectrum adopted in this study and examine its impact on results with a different claim. The spectrum of the primordial magnetic field in the observable $B_{0}$-$\lambda_{B}$ plane depends on the generation mechanism in the early universe \citep{Durrer:2013pga}. In general, for a magnetic field generated during the phase transition, its spectrum is blue with a positive power-law scaling $P_{B}\sim k^{2}$. As the universe expands, the field evolves on small scale and eventually relics in the pink region of $B_{0}$-$\lambda_{B}$ parameter space at present day. This conclusion holds for any blue spectrum generated before recombination due to the MHD process at recombination, including a typical scaling $P_{B}\sim k$ from inflationary magnetogenesis. However, the particular acausal property of inflation can in principle generate a red spectrum, for instance a (nearly) scale invariant one, which can fill the whole allowed parameter region depending on the onset of the magnetogenesis during inflation.

To explore the effect of different spectral models on the results, we do not take into account magnetogenesis as well as evolution dynamics, but instead simply assume the whole parameter space is filled by a single type of spectrum. Again we consider two typical cases: $P_{B}\sim k^{2}$ and $P_{B}\sim k^{-3}$, as in the main context. The numerical results are presented in Fig.  \ref{fig-peaked-All}. It demonstrates that a scale-invariant spectrum requires a relatively weaker magnetic field to simultaneously explain both radio anomalies compared to a peaked spectrum. In fact, this difference arises from the fact that the ALP-photon conversion probability in a stochastic magnetic background is higher for a scale-invariant spectrum than for a peaked one (see Fig. 4 in our previous work Ref. \citep{Addazi:2024kbq}).

\section*{Appendix F} 	

In the main body of the work we have assumed a frequency independent spectrum of the ALP for a simple consideration. In this appendix we generalize to a frequency dependent ALP spectrum and study its effects on the radio observations as well as the y-type distortion on CMB.  It is well known that at temperatures lower than $\mathcal{O}(10)$ eV, corresponding to $z\lesssim 10^5$, Compton scattering becomes inefficient at changing photon energy and transferring energy between photons and electrons, leading to a CMB y-distortion from the black-body spectrum via the Sunyaev-Zeldovich effect (see Ref. \cite{Hook:2023smg}  for more details). The y-distortion is significant suppressed at the end of recombination when the photons become free streaming cross the universe.  On the other hand, the CMB spectrum and its distortion is precisely measured in the frequency range from $60$GHz to $600$GHz from COBE \cite{Fixsen:1996nj}. Therefore, we estimate the y-type distortion induced by the excess photon converted from ALP in the y-era around $10^3\lesssim z \lesssim 10^5$ and in frequency range 10GHz$\lesssim f \lesssim$THz. 

Regarding the properties of the ALP, again we set  its mass $m_a=3 \times 10^{-13}$eV and photon-coupling constant $g_{a \gamma}=5 \times 10^{-13}\textrm{GeV}^{-1}$. Besides the constant ALP density spectrum $\Omega_{ALP}(\omega_0)=3 \times 10^{-6}$ investigated in the main body of our work, here we explore two other specific spectra: a red spectrum $\Omega_{ALP}(\omega_0)=10^{-6}(\omega_0/T_0)^{-0.5}$ with IR cutoff $\omega_0>10^2$Hz, and a blue spectrum  $\Omega_{ALP}(\omega_0)=10^{-5}(\omega_0/T_0)^{0.5}$ with UV cutoff $\omega_0<10^{18}$Hz. These cutoffs are chosen so as to saturate the bound of extra effective radiation constrain $\Omega_{ALP}/0.23\Omega_{\gamma}=\Delta N_\mathrm{eff}\lesssim 0.3$. Note that the scale invarinat spectrum is insensitive to both UV and IR cutoffs. 

For an order-of-magnitude estimation on y-type distortion, it suffices for us to  adopt the domain like model by assuming the constant magnetic field with strength $0.1$nGs and correlation length $1$Mpc. In one domain with size $d$ at certain redshift $z$, the conversion probability from ALP to photon is 
\begin{equation}
	\mathcal{P}(z)=\kappa^2 B^2(z) l_{osc}^2(z) \mathrm{sin}^2\left(\frac{d(z)}{l_{osc}(z)}\right). 
\end{equation}
The  oscillation length is $l_{osc}(z)=1/\sqrt{\Delta^2_{a\gamma}+\left(\Delta_{pl}+\Delta_{QED}+\Delta_{CMB}-\Delta_{a} \right)^2}$, where the plasma, QED and CMB effects are all taken into account. The  expression of these mass terms can be obtained from Ref. \cite{Schiavone:2021imu}. In the parameter space of our interests, the plasma effect term $\Delta_{pl}$ dominates over other terms, as shown in left panel in Fig.  \ref{fig-ytype}. Due to the dense plasma in y-era, the ALP-photon mixing can only be approximated as a closed system with conversion of the particle numbers within the mean free path of the photon. Indeed, from middle panel in Fig.  \ref{fig-ytype}, it is valid to choose $d(z)\simeq \lambda_\mathrm{MPF}$  because it is shorter than the correlation length of the magnetic field and the Hubble radius. Moreover, one can see that the mean free path length is much larger than the oscillation length in this relevant redshift region. Combined with the scaling $ \Delta_{pl} \propto z^2$ and $ B \propto z^2$, the conversion probability is actually a constant $\mathcal{P}=\kappa^2 B_0^2/\Delta_{pl0}^2$  during y-era. Thus the total conversion probability simplifies to  
$\mathcal{P}^{tot}=\sum_{i=1}^{n}\mathcal{P}(z_i)\simeq n \mathcal{P}(z) $, where the number of domains $n$ during $10^3\lesssim z \lesssim 10^5$ can be easily calculated as   $n \simeq 10^4$.

Given the conversion probability from ALP to photon, the relative distortion on the spectral intensity $\delta I/I$ can be obtained using formulas in Appendix B,
\begin{equation}
	\frac{\delta I}{I}=\frac{\pi^4}{15}\frac{e^x-1}{x^4}\frac{\Omega_{ALP}(x)}{\Omega_{\gamma}} \mathcal{P}^{tot}
\end{equation}    
with $x=\omega_0/T_0$. We apply this formula to three specific ALP spectra $\Omega_{ALP}\sim x^{-0.5,0,0.5}$ mentioned above. The constrains on y-type distortion of CMB is $|y|<1.5\times 10^{-5}$ with a certain shape as shown in right panel in Fig. \ref{fig-ytype} (see Refs. \cite{Fixsen:1996nj,Mukherjee:2018oeb,Hook:2023smg}). It demonstrates that the distortion induced by the ALP-driven excess radiation in our model is far below the CMB constrains, which also indicates a broad viable parameter space for other different scaling power of the ALP spectrum. In addition, we also explore the effect of  frequency dependent ALP spectrum on two radio abnormal signals in the RJ region of CMB. As shown in Fig. \ref{fig-ALPspect}, the red spectrum with proper scaling $\Omega_{ALP}\sim \omega^{-0.5}$ can produce a better fitting on excess radio data than the constant and blue spectra. Regarding the 21cm global signal, the red spectrum generates a deeper trough than the constant and blue spectra at $z\sim 17$, as expected.

\begin{figure}[tbp]			
	\includegraphics[scale=0.45]{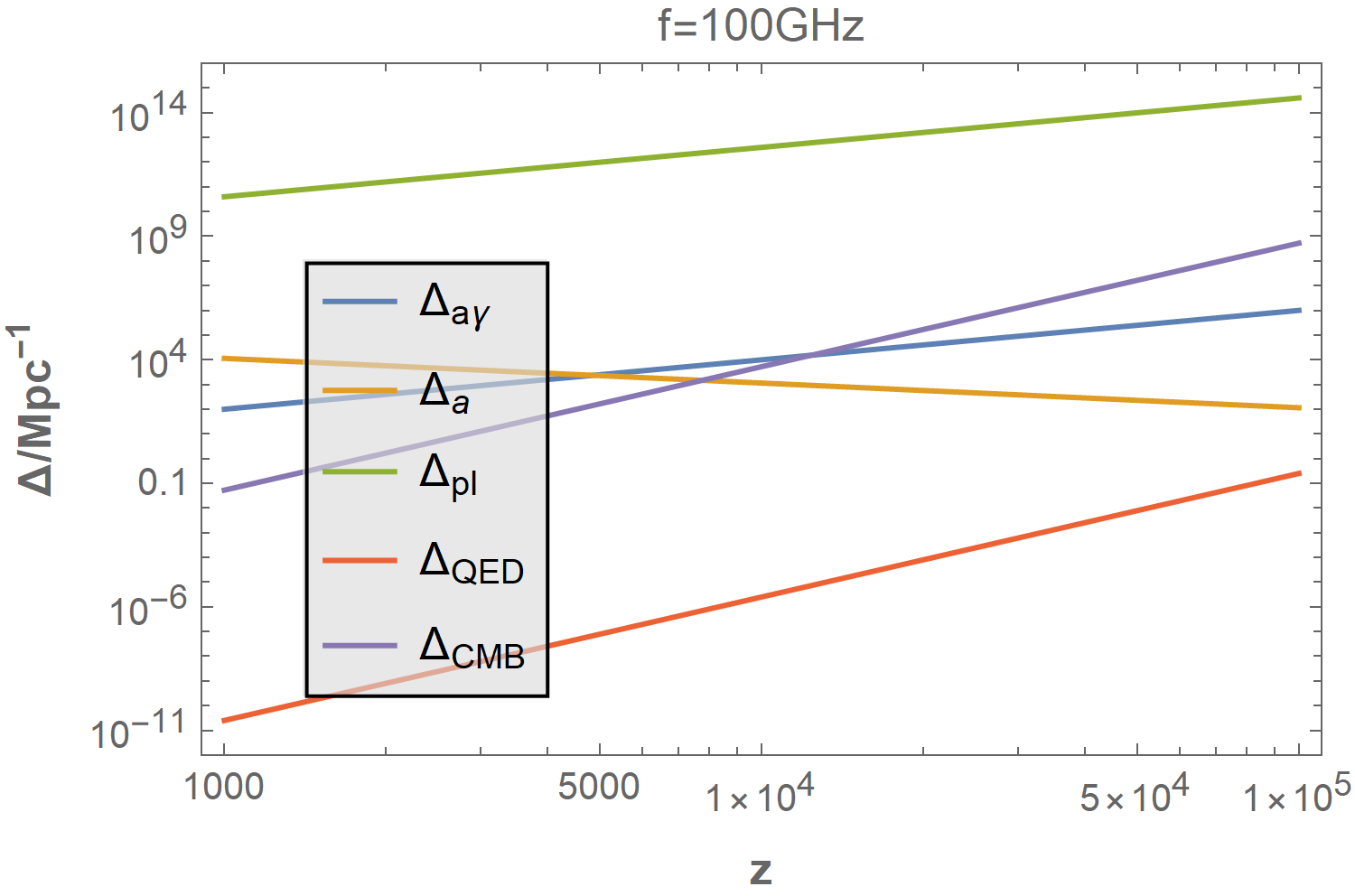} \includegraphics[scale=0.45]{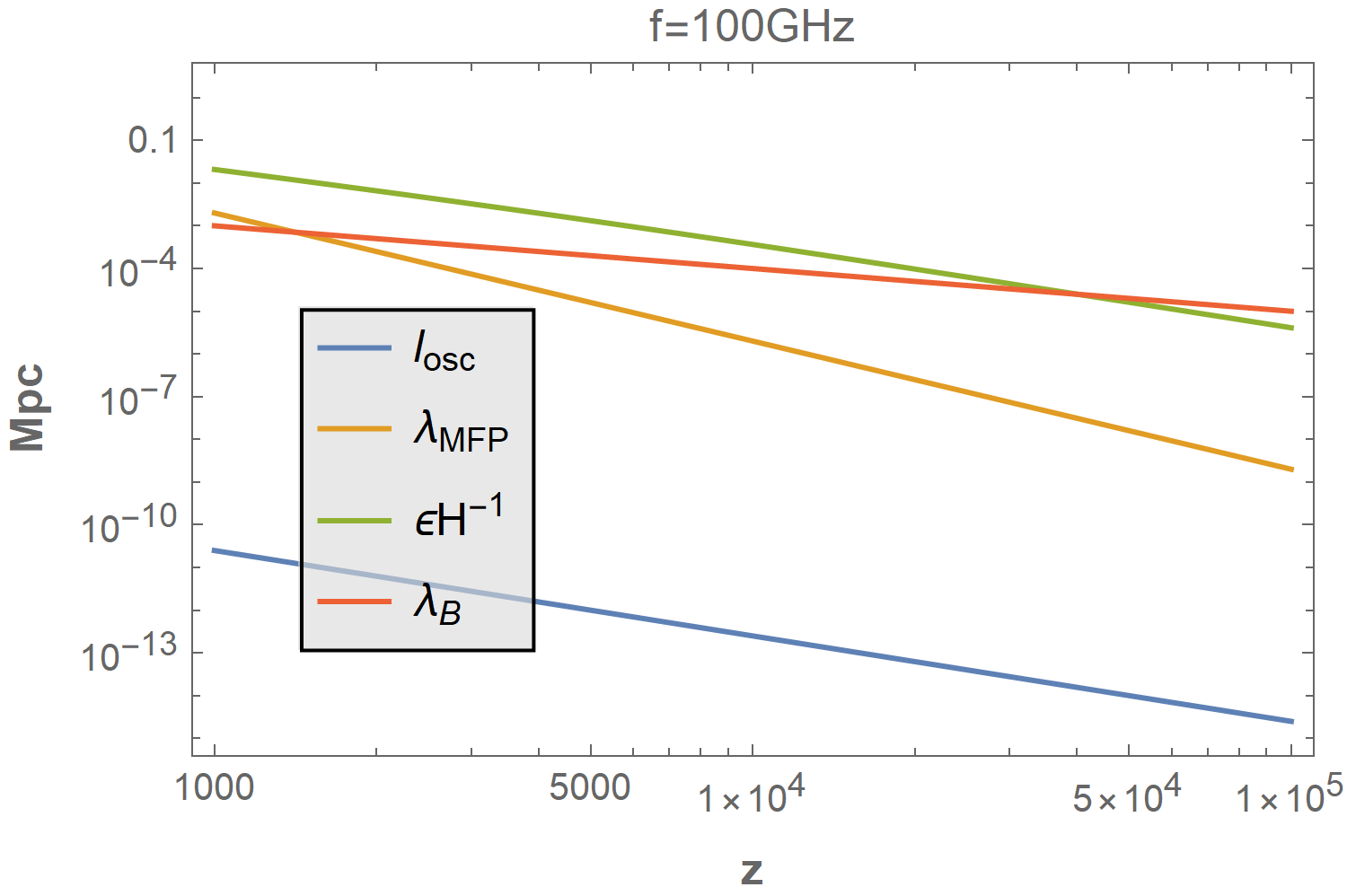}
	\includegraphics[scale=0.45]{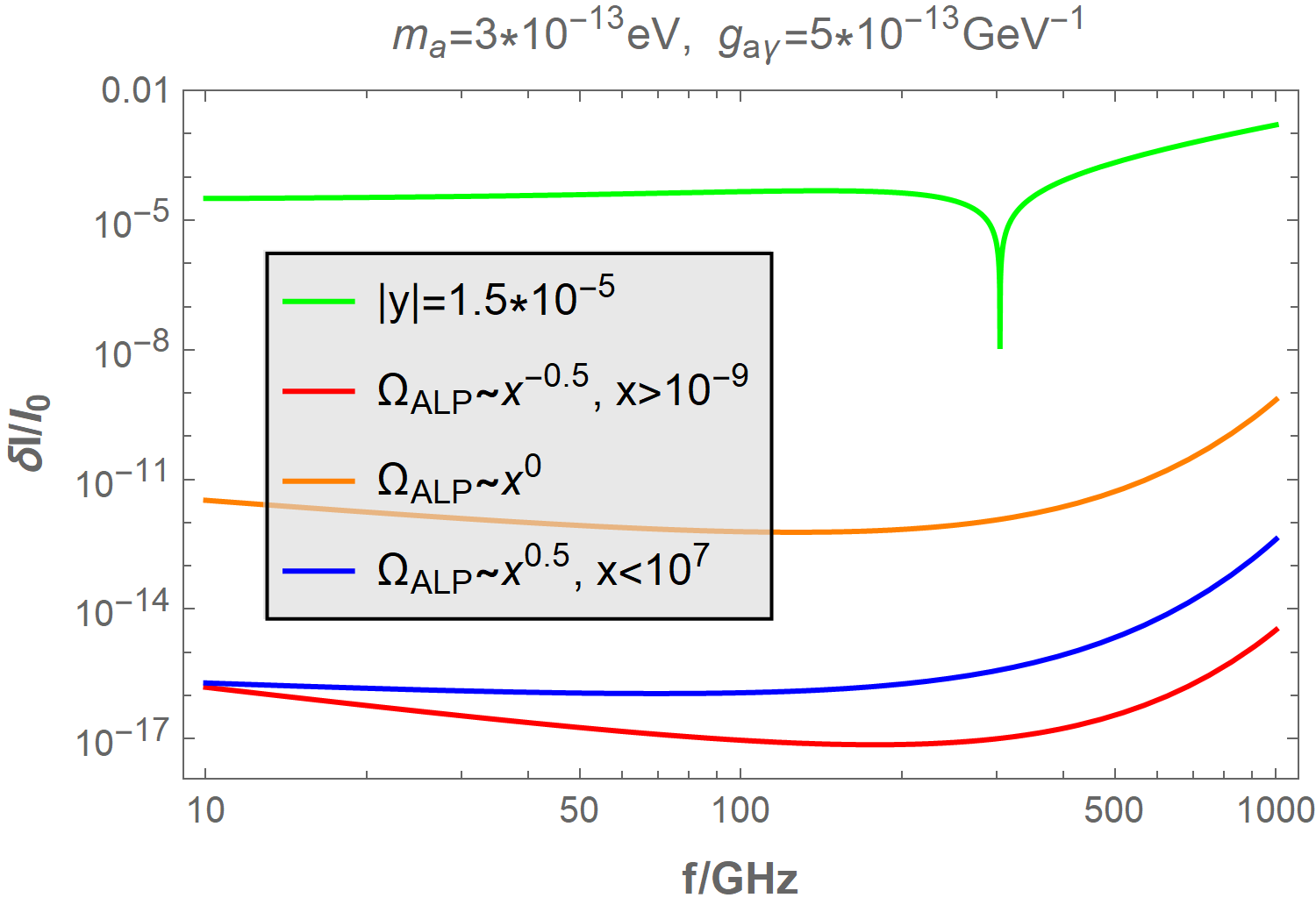}	
	
	\caption{{\bf Left:} Several mass terms in the mixing matrix during redshift $10^3<z<10^5$ at present day's frequency $f=100$GHz. The mass term $\Delta_{pl}$ denoting plasma effect dominates in this redshift range. {\bf Middle:} Comparison of typical length scales including the oscillation length of mixing, the mean free path of the photon,  correlation length of magnetic field and the characteristic scale of the steady approximation of expanding universe ($\epsilon=0.1$ same as the main body of work). {\bf Right:} y-type distortion induced by the ALP-photon conversion for three specific spectrum cases: a red spectrum $\Omega_{ALP}\sim \omega^{-0.5}$, a scale invariant spectrum $\Omega_{ALP}\sim const$ and a blue spectrum $\Omega_{ALP}\sim \omega^{0.5}$. The green line denotes y-type distortion bound from the CMB spectrum analysis within  frequency range from 10GHz to 1000GHz. In all three panels we set parameters $m_a=3 \times 10^{-13}$eV, $g_{a \gamma}=5 \times 10^{-13}\textrm{GeV}^{-1}$ and $B_0=0.1$nG.    }
	
	\label{fig-ytype}
\end{figure}

\begin{figure}[tbp]			
	\includegraphics[scale=0.6]{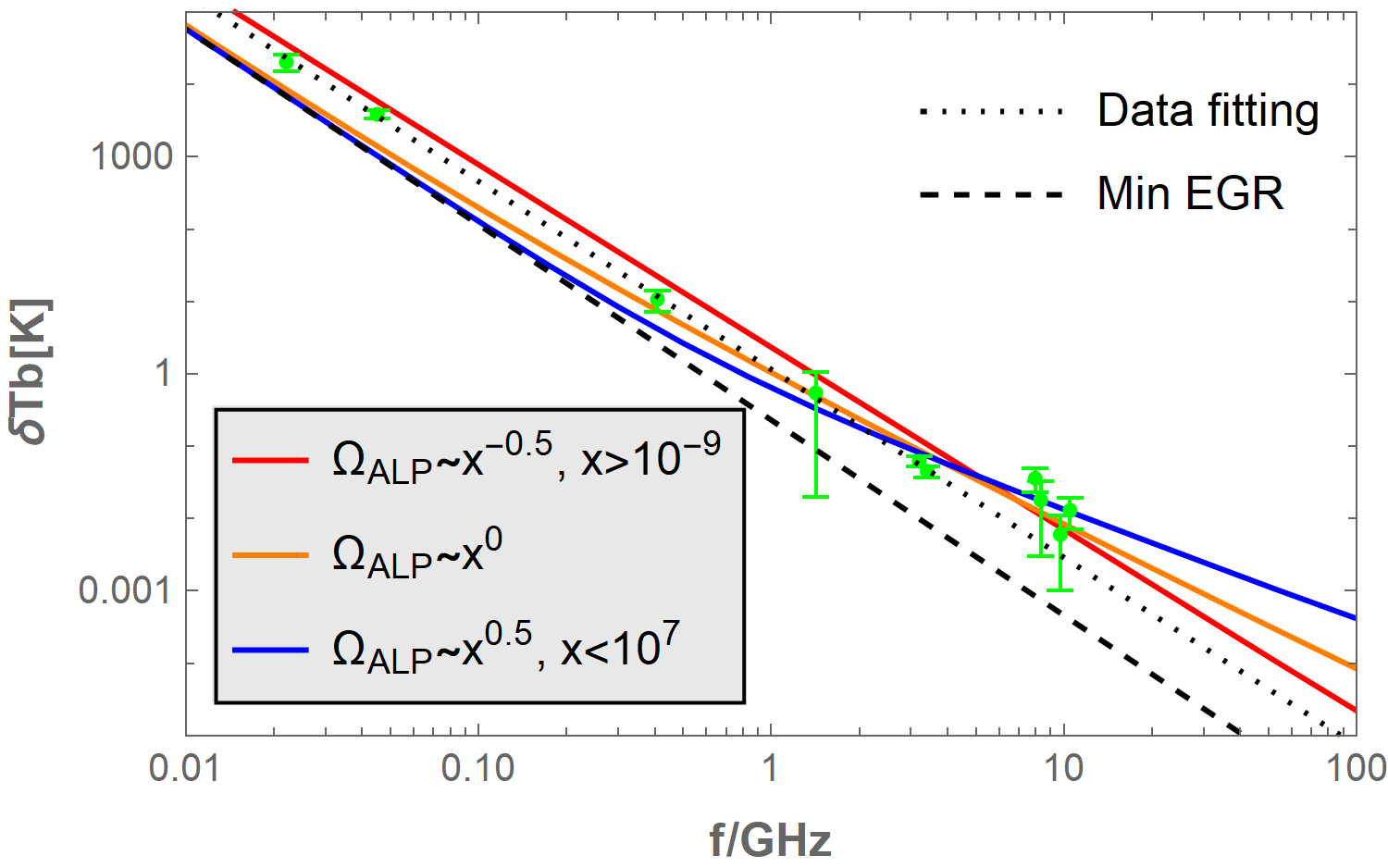} \includegraphics[scale=0.6]{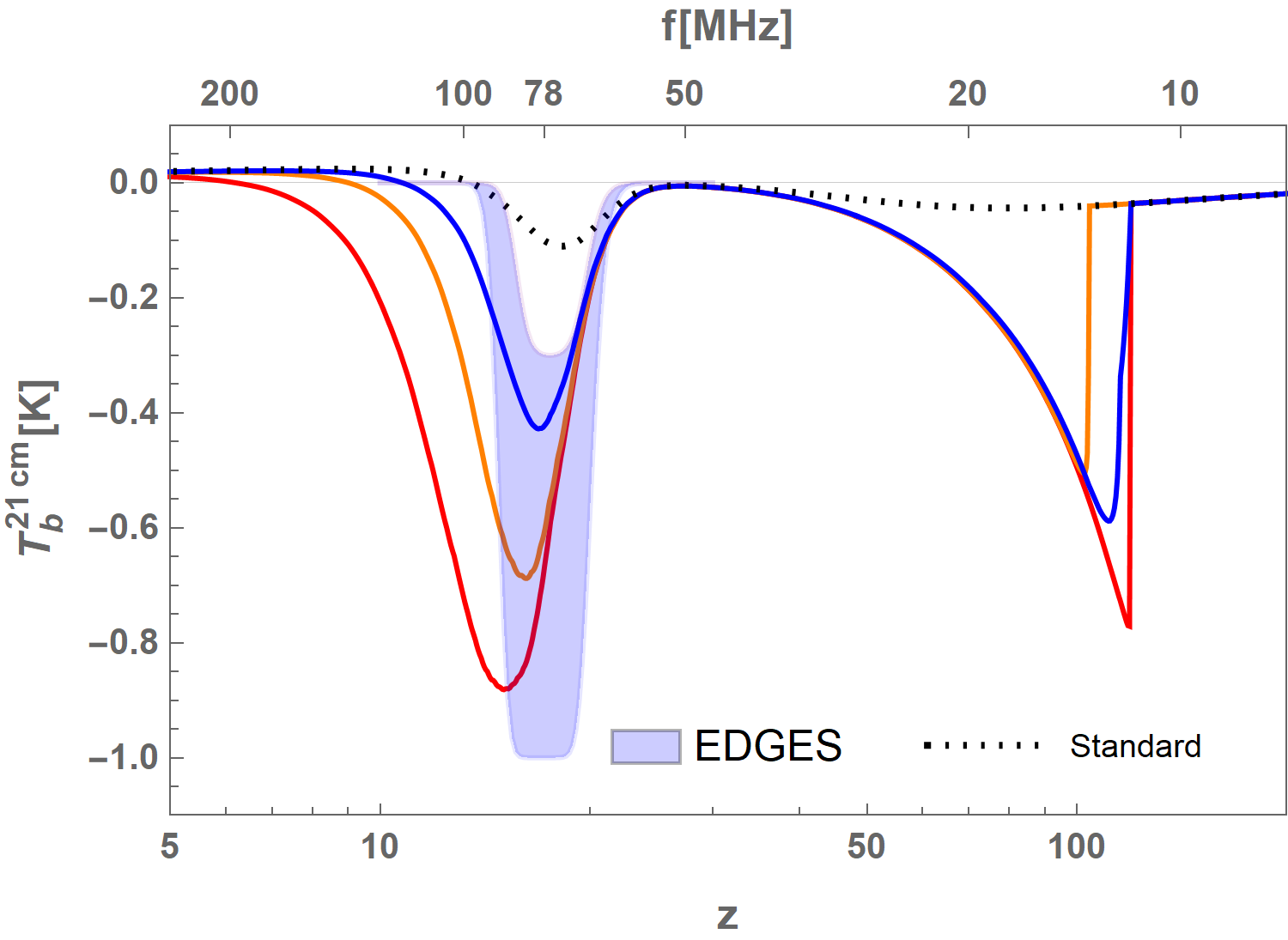}

	\caption{Fitting radio excess signal in the RJ region of CMB for three cases with blue, red and scale invariant ALP spectra (left) and their corresponding global 21cm lines (right). Parameters  are set to $m_a=3 \times 10^{-13}$eV, $g_{a \gamma}=5 \times 10^{-13}\textrm{GeV}^{-1}$ and $B_0=0.1$nG.  }
	
	\label{fig-ALPspect}
\end{figure}

\end{widetext}

\bibliography{Multimessenger}

\end{document}